\documentclass[iop,apj,tighten]{emulateapj}
\usepackage{amssymb}
\usepackage{amsmath}
\usepackage{times}
\usepackage{epsfig}
\usepackage{graphicx}
\usepackage{natbib}
\usepackage{color}
\usepackage{enumerate}
\include{def_bibtex}
\input{epsf}
\usepackage[T1]{fontenc}

\newcommand\sloanr{$r^\prime$}
\newcommand\sloani{$i^\prime$}

\newcommand\sloanri{$(r^{\prime} - i^{\prime})$}
\newcommand\Msol{M$_{\odot}$}

\begin{document}

\title{The Cheshire Cat Gravitational Lens: The Formation of a Massive Fossil Group}
\shortauthors{Irwin et al.}

\author{Jimmy A. Irwin\altaffilmark{1}, Renato Dupke\altaffilmark{1,2,3,4}, Eleazar R. Carrasco\altaffilmark{5},
W. Peter Maksym\altaffilmark{1}, Lucas Johnson\altaffilmark{1}, and Raymond E. White III\altaffilmark{1}}
%\makeauthors

\altaffiltext{1}{Department of Physics and Astronomy, University of
Alabama, Box 870324, Tuscaloosa, AL 35487, USA; e-mail: jairwin@ua.edu}

\altaffiltext{2}{Department of Astronomy, University of Michigan, 500 Church St., Ann Arbor, MI 48109, USA}

\altaffiltext{3}{Observat\'orio Nacional, Rua Gal.\ Jos\'e Cristino 77, S\~ao Crist\'ov\~ao, CEP20921-400 Rio de Janeiro RJ, Brazil}

\altaffiltext{4}{Eureka Scientific Inc., 2452 Delmer St. Suite 100, Oakland, CA 94602, USA}

\altaffiltext{5}{Gemini Observatory/AURA, Southern Operations Center, AURA, Casilla 603,
La Serena, Chile}

\submitted{Accepted for pubilcation in ApJ}

\begin{abstract}
The Cheshire Cat is a relatively poor group of galaxies dominated by two luminous elliptical galaxies surrounded by at least four
arcs from gravitationally lensed background galaxies that give the system a humorous appearance. Our combined optical/X-ray study
of this system reveals that it is experiencing a line of sight merger between two groups with a roughly
equal mass ratio with a relative velocity of $\sim$1350 km s$^{-1}$.
One group was most likely a low-mass fossil group, while the other group would have almost fit the classical definition of a fossil
group. The collision manifests itself in a bimodal galaxy velocity distribution, an elevated central X-ray temperature
and luminosity indicative of a shock, and gravitational arc centers that do not coincide with either large elliptical galaxy. One of the luminous
elliptical galaxies has a double nucleus embedded off-center in the stellar halo. The luminous ellipticals should merge
in less than a Gyr, after which observers will see a massive $1.2-1.5 \times 10^{14}$ M$_{\odot}$ fossil group with an $M_r = -24.0$ brightest group galaxy at its center.
Thus, the Cheshire Cat offers us the first opportunity to study a fossil group progenitor.
We discuss the limitations
of the classical definition of a fossil group in terms of magnitude gaps between the member galaxies.
We also suggest that if the merging of fossil (or near-fossil) groups is a common avenue for creating present-day fossil groups,
the time lag between the final galactic merging of the system and the onset of cooling in the shock-heated core could account
for the observed lack of well-developed cool cores in some fossil groups.
%The group already has a very high
%mass concentration characteristic of early assembly despite not formally being a fossil
%group yet, illustrating the limitations of the classical definition of a fossil group in terms
%of identifying groups that assembled a majority of their mass at very early times.
\end{abstract}

\keywords{gravitational lensing: strong, galaxies: clusters: individual: SDSS J1038+4849, X-rays: galaxies: clusters}

\section{INTRODUCTION}
\label{sec:intro}

Strong gravitational lensing of background galaxies by foreground groups or clusters of galaxies
is a powerful tool for determining the mass of the intervening system.
% in particular providing
%constraints on the presence of dark matter in merging clusters such as the Bullet Cluster (1E 0657-558;
%Clowe et al.\ 2006), and the Pandora Cluster (Abell 2744; Merten et al.\ 011) when combined with X-ray analysis.
Unlike other mass determination methods that require the assumption of hydrostatic
equilibrium of the cluster gas or dynamical assumptions about the constituent galaxies, the lensing method is
insensitive to the dynamical state of the lensing matter, and is dependent only on the amount of
mass between the source and the observer. In general, lensing mass estimates agree reasonably well
with the hydrostatic equilibrium method, at least for relaxed clusters \cite[e.g.,][]{wu00,rzepecki07,halkola08}.
Although X-ray/lensing joint analyses have been used mostly
for massive clusters, for which the lensing signal is stronger, its application for groups of galaxies has
become more robust \cite[e.g.,][]{grant04,fassnacht08}, providing a new window to study
group dynamics.

As the strength of lensing depends on the mass concentration of the lens, lens searches should
be most adept at finding lensing systems which, at a given mass, have the highest mass concentrations, such
as fossil groups of galaxies. Fossil groups are classically defined as systems dominated by a single giant
elliptical galaxy for which there is at least a two magnitude difference between that galaxy and the second rank galaxy (in $r$-band)
within 0.5~$r_{200}$ (where $r_{200}$ approximates the virial radius where the average mass density inside this radius is 200 times
the critical density of the Universe), and are bright sources of extended X-ray emission \cite[$L_{X,bol} >
10^{42} h_{50}^{-2}$ erg s$^{-1}$;][]{jones03}. Given the lack of bright galaxies other than the central brightest group galaxy (BGG), the correlation found
between concentration and formation epoch in $N$-body simulations \citep{wechsler02}, and the
fact that fossil groups seem to have higher central mass concentrations than non-fossil groups
(\citealp[e.g.,][]{khosroshahi07}; \citealp[c.f.,][]{sun09}), it was suggested that fossil groups formed earlier than normal
groups, with most of the larger galaxies
merging into a single central dominant elliptical galaxy, with the system then remaining undisturbed for a very long time
until the present day \citep{ponman94,vikhlinin99,jones00}. In this picture, age and a lack of
subsequent interaction with another group would be the key differences between a fossil group and a normal group.

However, more recently, X-ray and optical measurements of fossil groups have shown several potential inconsistencies with
this formation mechanism. Fossil groups appear to resemble clusters of galaxies more closely than groups of
galaxies despite their low galaxy count \cite[e.g.,][]{proctor11}. The most massive fossil groups have a hot intergalactic
medium similar to that of clusters, sometimes in excess of 4 keV \cite[e.g.,][]{khosroshahi06}.
Measurements of galaxy velocity dispersions in fossil groups \citep{cypriano06, mendes06, miller12}
seem to be consistent with the measured $T_X$, at least for the few fossil groups with
relatively good X-ray data. This is further suggested by their (not atypical) location in the
$L_X-T_X$ relation \cite[e.g.,][]{khosroshahi07,miller12}. This would suggest that they
have relatively deep gravitational potential wells typical of clusters, and not normal groups.
Other studies, however, have not found substantial differences in the X-ray and optical scaling relations of
fossil groups and normal groups \citep{sun09,voevodkin10,harrison12,girardi14}.
A further inconsistency is that despite having high measured mass concentrations which would indicate early formation
epochs, fossil groups in general do not have well-developed cool cores as would be expected from an old system with a cooling time
less than the Hubble time \citep{khosroshahi04,khosroshahi06,sun04}. Conversely, non-fossil groups often do exhibit
cool cores \citep{finoguenov99}.

Cosmological simulations indicate that the fossil stage of a group might be transitory in nature over
the lifetime of the group. $N$-body simulations of a cube 80 Mpc on a side by \citet{vonbendabeckmann08} showed that
nearly half of the simulated fossil groups identified at $z$=0 would not be identified as such at $z$=0.3, and virtually none
would be identified as fossil groups at $z$=0.93. Conversely, few of the fossil groups identified at $z$=0.93 remained fossil groups until $z$=0,
with subsequent infall of one or more luminous galaxies after $z$=0.93 removing the group from fossil group status by dropping
the magnitude gap between the first and second rank galaxies below 2.0. In addition, \citet{dariush10} find that
90\% of fossil groups identified in simulations by magnitude gap arguments become non-fossil groups after $\sim$4 Gyr as a result of infall
of luminous galaxies. Thus,
it is not clear whether fossil groups truly have a different formation mechanism from normal groups. On one hand, their
observed galaxy luminosity distributions
% and hot gas properties
are distinct from non-fossil groups; on the other
hand the fossil group status of a group seems rather easy to alter with simply the infall of a single luminous galaxy, a process
that apparently occurs frequently according to simulations.

In this paper, we use the strongly lensed Cheshire Cat galaxy group to better understand how groups pass in and out of
the fossil group stage via merging.
We assume a cosmology of $H_0$ = 70 km s$^{-1}$ Mpc$^{-1}$,  $\Omega_{M} = 0.27$,
and  $\Omega_{\Lambda} = 0.73$, so that at the group redshift of $z=0.431$, 1$^{\prime\prime}$ = 5.67 kpc.
All our masses and radii are given in units of $(h/h_{70})^{-1}$, while all our equations specifically state
the $h_{70}^{-1}$ dependence.
We discuss the properties of the group in \S~\ref{sec:cheshire_cat}. We describe the {\it Gemini} GMOS, {\it Chandra}, and
{\it Hubble Space Telescope (HST)} data preparation and analysis in \S~\ref{sec:gemini}, \S~\ref{sec:xray}, and \S~\ref{sec:hst}, respectively.
We discuss our results in \S~\ref{sec:discuss}, and summarize our findings in \S~\ref{sec:summary}.

\section{The Cheshire Cat Gravitational Lens}
\label{sec:cheshire_cat}

\begin{figure*}
\hspace{20mm}
 \includegraphics[height=11.5cm,width=14.5cm]{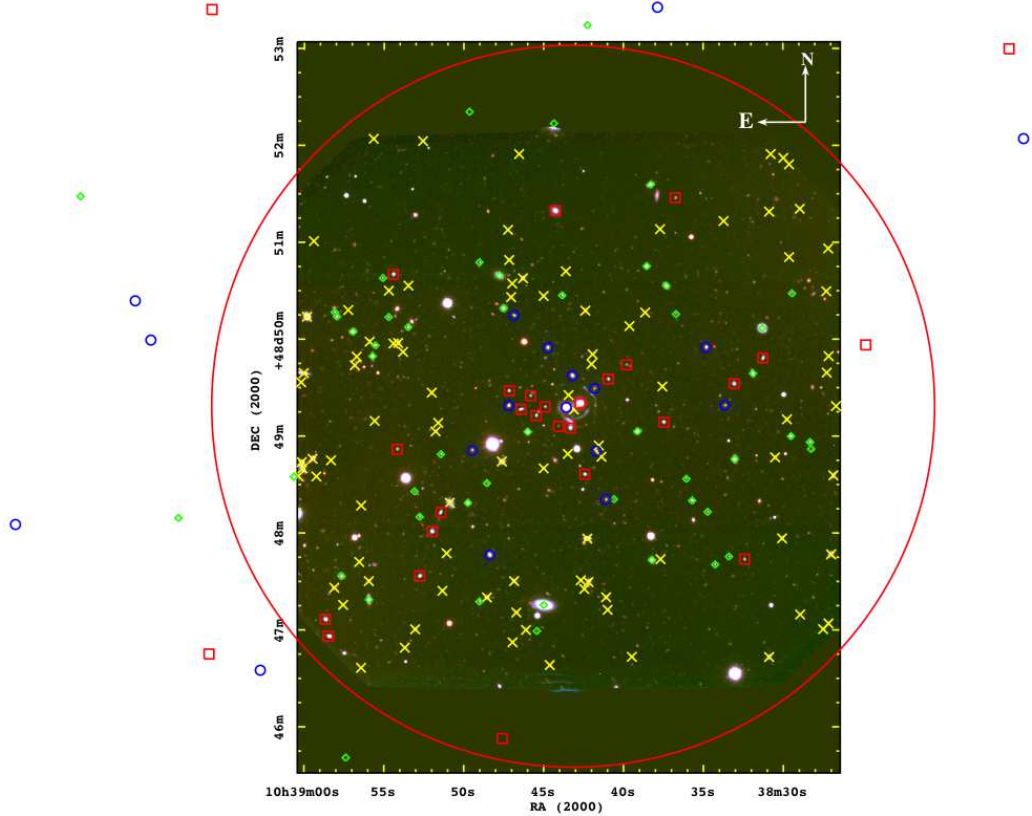}
\vspace{-3mm}
 \caption{GMOS-S pseudo-color image of the Cheshire Cat group. The size of the image is $\approx 5.4 \times 5.4$
arcmin$^2$. Open (blue) circles are galaxies members of the G1 group and the open (red) squares are galaxies members
of the G2 group. Open (green) diamonds are background and foreground galaxies. Galaxies with no redshift information
and with \sloanr$< 23$ mag and \sloanr - \sloani$<0.8$ are represented by the (yellow) crosses. The red circle represents the virial
radius of 1.3 Mpc ($229^{\prime\prime}$).}\label{fig:spatialdist}
 \label{fig:coord}
\end{figure*}

As mentioned above, groups with high central mass concentrations should make them more
efficient strong gravitational lenses, capable of being detected by the Sloan Digital Sky Survey (SDSS). One such
very peculiar strong lens system is SDSS J103842.59+484917.7, informally known as the Cheshire Cat \citep{carroll65}
Lens. The Cheshire Cat Lens was discovered by the Cambridge Sloan Survey of Wide Arcs in the Sky
\citep{belokurov09}, and independently by \citet{kubo09}. The unusual visual nature of the
system (Figure~\ref{fig:spatialdist} and Figure~\ref{fig:chandra}) results from four separate background galaxies with redshifts ranging from 0.80 to 2.78
\citep{bayliss11} being lensed by a foreground galaxy group dominated by two closely spaced elliptical
galaxies. These two bright elliptical galaxies representing the ``eyes" of the Cheshire Cat are SDSS
J103843.58+484917.7 and SDSS J103842.68+484920.2 at redshifts $z=0.426$ and $z=0.433$,
respectively, surrounded by a collection of much fainter galaxies.
Hereafter, we will refer to SDSS J103843.58+484917.7 as the eastern eye and SDSS J103842.68+484920.2
as the western eye.
% The Cat's ``nose" is a fortuitously placed, but unrelated, foreground galaxy.
The Einstein radii of the four
lensed galaxies range from 9\farcs0 -- 12\farcs.5 (51--71 kpc).
The total projected mass enclosed within the outermost ring
is $3.3 \times 10^{13}$ M$_{\odot}$ \citep{belokurov09}, about a factor of ten greater than
the mass of the Local Group in a volume with a radius only slightly greater than the Milky Way--Large Magellanic Cloud
distance.
%Obviously this system fails the classical definition of a fossil group by the fact that
%there are two bright elliptical galaxies in the system rather than one, but once these ``eye" galaxies eventually
%merge (\S~\ref{ssec:future}), there will be a two magnitude gap between this merged galaxy and the next brightest
%galaxy in the system, after which the system will qualify as a fossil group.
The redshift difference between the two bright eye galaxies corresponds to
an 1800 km s$^{-1}$ rest frame velocity difference. We argue below that the two galaxies are at the same distance but with high
relative velocity, rather than the redshift difference being due to the Hubble flow. The high relative velocity combined
with other merging indicators presented below suggests that the eye galaxies are BGGs of two groups (which we designate
`G1' for the eastern eye group, and `G2' for the western eye group) that are in the
process of merging in a line of sight collision. Furthermore, our analysis
indicates that one group was most likely a fossil group prior to the merger, while the other group was a near-fossil group,
and that once
these eye galaxies merge, the system will become a massive fossil group. Thus, the
Cheshire Cat appears to be a prime example of fossil groups leaving, and eventually re-establishing their fossil group status.
Figure~\ref{fig:spatialdist} shows the spatial distribution of the spectroscopically confirmed galaxies members of the
Cheshire Cat group. Open (blue) circles are galaxy members of the G1 group and open (red) squares are the galaxy members
of the G2 group. The diamonds (yellow) are the spectroscopically confirmed background and foreground galaxies. Galaxies
with no redshift information are marked with a cross (yellow).

Previous work has studied the galaxy population and strong/weak lensing of the Cheshire Cat and established a virial mass ($M_{200}$) in the
range of 1.0--2.9 $\times 10^{14}$ M$_{\odot}$, $r_{200}$ of 1.5 Mpc, a concentration parameter ($c_{200}$)
of 17--34, and a galaxy richness parameter ($N_{200})$ of 25 \citep[all values have been scaled to our adopted cosmology;][]{bayliss11,oguri12,wiesner12}.
In particular, $c_{200}$ is quite high compared to other lensed systems studied in
\cite{oguri12} and \cite{wiesner12}.

\section{{\it Gemini} and Imaging and Spectroscopic Data Analysis}
\label{sec:gemini}

The optical observations (imaging and spectroscopy) were performed with the {\it Gemini} Multi-Object
Spectrograph \cite[hereafter GMOS;][]{hook04} at the {\it Gemini} North Telescope in Hawaii, in queue
mode, as part of the program GN-2011A-Q-25.

The direct images were recorded through the \sloanr~and \sloani~filters during the night of
2011 January 4, in dark  time, with seeing median values of 0\farcs8 and 0\farcs9 for the \sloanr~and \sloani~filters,
respectively. The night was not photometric. Three 300 s exposures (binned by two in both axes,
with pixel scale of 0\farcs146)  were observed in each filter. Offsets between  exposures were used to
take into account the gaps between the CCDs (37 un-binned pixels) and for cosmic ray removal. The
images were processed with the {\it Gemini} GMOS package version 1.8 inside
IRAF\footnote[6]{IRAF is distributed by NOAO, which is operated by the Association of Universities for
Research in Astronomy Inc., under cooperative a agreement with the National Science Foundation.}. The
images were bias/overscan-subtracted, trimmed and flat-fielded. The final processed images were
registered to a common pixel position and then combined.  Because the group was observed under non-photometric
conditions, we used the values listed in the Gemini Performance Monitoring
webpage\footnote[7]{http://www.gemini.edu/sciops/instruments/performance-monitoring/dat a-products/gmos-n-and-s/photometric-zero-points}
of 28.22 and 28.15 mag for \sloanr~and \sloani~filters, respectively (values
closest to the date of the observations), as an approximate zero point for the photometry.

Object detection and extraction of the photometric parameters were obtained with the program
{\tt Source Extractor} \citep{bertin96}.
The parameter MAG$\_$AUTO was adopted as the value for the total magnitude of the objects. Colors
were derived by measuring fluxes inside a fixed circular aperture of 1\farcs2 in diameter ($\sim$8 pixels)
in both filters. Because the images were observed under non-photometric conditions, we have to determine
the correction (offset) necessary to apply to the magnitudes due to the effect of the extinction produced
by the clouds during the observations.  The correction was determined by comparing the
magnitudes of well isolated and non-saturated field stars, measured at a fixed circular aperture of
3\arcsec, with the magnitudes of the same stars, measured inside the same aperture (the fiber
magnitude), from the Sloan Digital Sky Survey Data Release 8 (SDSS DR8)\footnote[8]{http://
www.sdss3.org/dr8/}. The comparison yielded an offset between the SExtractor magnitudes and the
SDSS DR8 magnitudes of $-0.41\pm0.07$ mag and $-0.41\pm0.12$ mag for the \sloanr~and
\sloani~filters, respectively. The magnitudes in the final catalog were corrected by these offsets.
The final galaxy catalog contains the total magnitudes in \sloanr~ and \sloani, colors and other structural
parameters for 595 objects classified as galaxies, inside an area of $\sim 5 \times 5$ arcmin$^2$. The
galaxies were defined as objects with SExtractor {\em stellarity} flag $\le0.6$ and brighter than \sloanr
 = 24.5 mag ($M_{r^{\prime}}=-17.7$ mag), which is the magnitude where the number counts start to turn
over.

All galaxies brighter than \sloanr $=$ 22 mag and with \sloanr - \sloani $< 0.8$ mag were selected for spectroscopic
follow-up (151 galaxies). Two masks were designed for the observations, one with 36 galaxies and another
with 39 galaxies ($\sim$50\% of the selected sample). The two masks were observed between 2011
May 23 and 2011 July 01. All spectra were observed during dark time, with median seeing values
of 0\farcs8 and under photometric conditions. The spectra were acquired using the R400 grating
centered at 7000~\AA, in order to maximize the wavelength coverage for galaxies at the cluster
distance. For each mask, a total exposure time of 2.5 hr ($5 \times 1800$ s) was used. Small
offsets of $\sim$50~\AA~in spectral direction toward the blue and/or the red were applied between
exposures to avoid any lost or important emission/absorption lines presented in the spectra that could
lie in the gaps between the GMOS CCDs. Spectroscopic dome flats and CuAr comparison lamp spectra were
taken before or after each science exposure.

The spectra were reduced with the {\it Gemini} GMOS package, following the standard procedure. The science
spectra, dome flats and comparison lamps were bias-subtracted and trimmed. Spectroscopic dome flats were
processed by removing the calibration unit plus GMOS spectral response and the calibration unit uneven
illumination, normalizing and leaving only the pixel-to-pixel variations and the fringing. The 2-D spectra were
then flat-fielded, wavelength calibrated (rms $\sim$ $0.15 - 0.20$~\AA), corrected by S-shape
distortions, sky-subtracted and extracted  to a one-dimensional format using a fixed aperture of
1\farcs2 ($1.5 \times$ FWHM).  The final extracted spectra have a resolution of $\sim$$7.5$ \AA\ at
$\sim$$7000$ \AA\ (measured from the arc lines FWHM) with a dispersion of $\sim$$1.37$ \AA\
pixel$^{-1}$, covering a wavelength interval of $\sim$$4000$--8400~\AA\ (the wavelength coverage
depends on the position of the slit in the GMOS field of view).

The redshifts of the galaxies were determined using the cross-correlation technique or emission line
fitting.  For galaxies with clear emission lines, the routine RVIDLINE in the IRAF RV
package was used employing a line-by-line gaussian fit to measure the radial velocity. The residual
of the average velocity shifts of all measurements were used to estimate the errors. For early-type
galaxies, the observed spectra were cross-correlated with high signal-to-noise templates using the
FXCOR program in the RV package inside IRAF. The errors given by FXCOR were estimated using the $R$
statistic of \citet{tonry79}.  

We were able to measure the redshift for 75 galaxies (100\% success rate) of which 34 had redshifts consistent
with the Cheshire Cat. The calculated redshift,
corrected to the heliocentric reference frame, the corresponding errors, the magnitudes and other
relevant information are presented in Table~\ref{tbl:tabgal}. To supplement our sample, spectroscopic redshifts of 
galaxies in the area of the Cheshire Cat from the work of \citet{bayliss14} were added to our sample.
Based on a spectroscopic sample of 168 galaxies, \citet{bayliss14}
obtained 23 redshifts within a physical radius of 2.0 Mpc of the center of the group that were not already in our GMOS catalog,
of which 14 were identified as Cheshire Cat galaxies.
%Three galaxies from their catalog are inside the GMOS field of view ($\sim$1.0 Mpc). Two of them are at the redshift of
%the group and one is a foreground galaxy.
The coordinates, magnitude and redshift of these 14 additional Cheshire Cat galaxies
and nine foreground/background galaxies are shown in Table~\ref{tbl:tabgal}, and labeled as B1--B23.

\begin{deluxetable*}{lccccccccl}
\tabletypesize{\scriptsize}
\tablecaption{Redshifts and magnitudes of the observed galaxies}
%\tablenum{1}
%\tablecolumns{10}
%\tablewidth{0pc}
\tablehead{
\colhead{ID} &
\colhead{R.A.} &
\colhead{Decl.} &
\colhead{\sloanr} &
\colhead{\sloani} &
\colhead{\sloanri} &
\colhead{$z$} &
\colhead{$\Delta (z)$} &
\colhead{$R$/\#em. lines} &
\colhead{Member?} \\
\colhead{(1)} &
\colhead{(2)} &
\colhead{(3)} &
\colhead{(4)} &
\colhead{(5)} &
\colhead{(6)} &
\colhead{(7)} &
\colhead{(8)} &
\colhead{(9)} &
\colhead{(10)}}
\startdata
  \bf 387 & \bf 10 38 31.884 & \bf $+$48 49 38.75 & \bf 20.76 & \bf 20.42 &  \bf 0.40 & \bf 0.246811 & \bf 0.000120 &\bf \nodata/~~~~~7 & \bf N   \\
  \bf 405 & \bf 10 38 33.036 & \bf $+$48 48 45.68 & \bf 20.01 & \bf 19.72 & \bf 0.33 & \bf 0.246617 & \bf 0.000267 &  \bf 3.83/\nodata & \bf N    \\
  \bf 403 & \bf 10 38 33.072 & \bf $+$48 49 32.38 & \bf 21.19 & \bf 20.51 &  \bf 0.67 & \bf 0.432056 & \bf 0.000133 &  \bf 4.79/\nodata & \bf Y/G2   \\
  \bf 446 & \bf 10 38 33.648 & \bf $+$48 49 19.02 & \bf 22.33 & \bf 21.70 &  \bf 0.59 & \bf 0.426248 & \bf 0.000087 &  \bf 7.99/\nodata & \bf Y/G1   \\
  \bf 532 & \bf 10 38 34.728 & \bf $+$48 48 13.10 & \bf 22.54 & \bf 21.74 &  \bf 0.78 & \bf 0.802925 & \bf 0.000110 &\bf \nodata/~~~~~2 & \bf N   \\
  \bf 535 & \bf 10 38 34.836 & \bf $+$48 49 54.95 & \bf 21.60 & \bf 20.90 &  \bf 0.66 & \bf 0.429267 & \bf 0.000177 &  \bf 4.52/\nodata & \bf Y/G1   \\
  \bf 590 & \bf 10 38 35.700 & \bf $+$48 48 20.09 & \bf 21.77 & \bf 21.36 &  \bf 0.38 & \bf 0.470159 & \bf 0.000083 &\bf \nodata/~~~~~6 & \bf N  \\
  \bf 611 & \bf 10 38 36.060 & \bf $+$48 48 33.55 & \bf 22.45 & \bf 21.75 &  \bf 0.73 & \bf 0.512468 & \bf 0.000077 &  \bf 8.97/\nodata & \bf N   \\
  \bf 654 & \bf 10 38 36.708 & \bf $+$48 50 15.54 & \bf 22.83 & \bf 22.39 &  \bf 0.40 & \bf 0.601393 & \bf 0.000267 &  \bf 3.77/\nodata & \bf N   \\
  \bf 679 & \bf 10 38 37.320 & \bf $+$48 50 33.11 & \bf 20.37 & \bf 20.05 &  \bf 0.39 & \bf 0.247048 & \bf 0.000267 &  \bf 4.89/\nodata & \bf N   \\
  \bf 702 & \bf 10 38 37.464 & \bf $+$48 49 08.69 & \bf 21.45 & \bf 20.80 &  \bf 0.64 & \bf 0.430621 & \bf 0.000100 &  \bf 9.56/\nodata & \bf Y/G2   \\
 \bf 694 & \bf 10 38 38.220 & \bf $+$48 47 43.33 & \bf 20.36 & \bf 19.81 &  \bf 0.57 & \bf 0.441562 & \bf 0.000267 &  \bf 5.16/\nodata & \bf N   \\
  \bf 764 & \bf 10 38 38.544 & \bf $+$48 50 45.20 & \bf 20.37 & \bf 20.04 &  \bf 0.34 & \bf 0.321436 & \bf 0.000267 &  \bf 4.95/\nodata & \bf N   \\
  \bf 788 & \bf 10 38 39.120 & \bf $+$48 49 02.96 & \bf 20.15 & \bf 19.73 &  \bf 0.42 & \bf 0.239652 & \bf 0.000043 &\bf \nodata/~~~~~8 & \bf N  \\
  \bf 830 & \bf 10 38 39.804 & \bf $+$48 49 44.22 & \bf 22.76 & \bf 22.12 &  \bf 0.58 & \bf 0.433140 & \bf 0.000153 &  \bf 5.20/\nodata & \bf Y/G2  \\
  \bf 893 & \bf 10 38 40.596 & \bf $+$48 48 20.95 & \bf 22.63 & \bf 22.03 &  \bf 0.59 & \bf 0.661674 & \bf 0.000254 &\bf \nodata/~~~~~4 & \bf N  \\
  \bf 905 & \bf 10 38 40.956 & \bf $+$48 49 35.26 & \bf 21.92 & \bf 21.30 &  \bf 0.64 & \bf 0.431102 & \bf 0.000103 & \bf 11.46/\nodata & \bf Y/G2  \\
  \bf 894 & \bf 10 38 41.100 & \bf $+$48 48 20.81 & \bf 22.19 & \bf 21.55 &  \bf 0.60 & \bf 0.426418 & \bf 0.000267 &  \bf 8.50/~~~~~4  & \bf Y/G1  \\
  \bf 952 & \bf 10 38 41.712 & \bf $+$48 48 50.72 & \bf 21.92 & \bf 21.26 &  \bf 0.66 & \bf 0.426842 & \bf 0.000063 & \bf 11.40/\nodata & \bf Y/G1  \\
  \bf 970 & \bf 10 38 41.820 & \bf $+$48 49 29.53 & \bf 22.30 & \bf 21.67 &  \bf 0.59 & \bf 0.426719 & \bf 0.000143 & \bf  8.23/\nodata & \bf Y/G1  \\
  \bf 1006 & \bf 10 38 42.432 & \bf $+$48 48 36.36 & \bf 21.57 & \bf 20.90 &  \bf 0.66 & \bf 0.433954 & \bf 0.000087 & \bf 12.14/\nodata & \bf Y/G2 \\
  \bf 1092 & \bf 10 38 42.720 & \bf $+$48 49 20.28 & \bf 19.15 & \bf 18.49 &  \bf 0.68 & \bf 0.432519 & \bf 0.000157 &  \bf 8.45/\nodata & \bf Y/G2 (W)  \\
 \bf 1012 & \bf 10 38 43.188 & \bf $+$48 49 37.45 & \bf 21.27 & \bf 20.64 &  \bf 0.65 & \bf 0.428253 & \bf 0.000160 & \bf  9.02/\nodata & \bf Y/G1  \\
  \bf 983 & \bf 10 38 43.332 & \bf $+$48 49 05.12 & \bf 20.98 & \bf 20.25 &  \bf 0.71 & \bf 0.433964 & \bf 0.000130 & \bf  8.15/\nodata & \bf Y/G2  \\
  \bf 972 & \bf 10 38 43.584 & \bf $+$48 49 17.83 & \bf 18.86 & \bf 18.21 &  \bf 0.67 & \bf 0.425845 & \bf 0.000140 & \bf  8.56/\nodata & \bf Y/G1 (E) \\
 \bf 1090 & \bf 10 38 43.836 & \bf $+$48 50 27.13 & \bf 22.04 & \bf 21.34 &  \bf 0.72 & \bf 0.552115 & \bf 0.000267 & \bf  4.84/\nodata & \bf N  \\
 \bf 1224 (B1)  & \bf 10 38 44.068 & \bf $+$48 49 06.11 & \bf 23.28 & \bf 22.67 &  \bf 0.61 & \bf 0.433700 & \bf 0.000600 &\bf \nodata        & \bf Y/G2  \\
 \bf 1124 & \bf 10 38 44.736 & \bf $+$48 49 54.70 & \bf 21.49 & \bf 20.87 &  \bf 0.64 & \bf 0.428783 & \bf 0.000093 & \bf 12.40/\nodata & \bf Y/G1  \\
 \bf 1164 & \bf 10 38 44.916 & \bf $+$48 49 18.12 & \bf 22.53 & \bf 21.94 &  \bf 0.62 & \bf 0.430498 & \bf 0.000093 &  \bf 8.92/\nodata & \bf Y/G2  \\
 \bf 1195 & \bf 10 38 45.456 & \bf $+$48 49 12.79 & \bf 21.96 & \bf 21.30 &  \bf 0.67 & \bf 0.433063 & \bf 0.000100 &  \bf 9.66/\nodata & \bf Y/G2  \\
 \bf 1222 & \bf 10 38 45.816 & \bf $+$48 49 24.96 & \bf 22.56 & \bf 21.94 &  \bf 0.66 & \bf 0.433743 & \bf 0.000127 &  \bf 7.33/\nodata & \bf Y/G2  \\
 \bf 1230 & \bf 10 38 45.996 & \bf $+$48 49 02.64 & \bf 20.24 & \bf 19.94 &  \bf 0.33 & \bf 0.378415 & \bf 0.000050 &\bf \nodata/~~~~~7 & \bf N  \\
 \bf 1265 & \bf 10 38 46.428 & \bf $+$48 49 16.64 & \bf 21.51 & \bf 20.82 &  \bf 0.66 & \bf 0.431549 & \bf 0.000110 &  \bf 8.06/\nodata & \bf Y/G2   \\
 \bf 1305 & \bf 10 38 46.860 & \bf $+$48 50 14.68 & \bf 22.01 & \bf 21.38 &  \bf 0.59 & \bf 0.429240 & \bf 0.000173 &  \bf 5.60/\nodata & \bf Y/G1   \\
 \bf 1329 & \bf 10 38 47.148 & \bf $+$48 49 28.09 & \bf 22.07 & \bf 21.38 &  \bf 0.66 & \bf 0.434998 & \bf 0.000147 &  \bf 7.09/\nodata & \bf Y/G2   \\
 \bf 1331 & \bf 10 38 47.184 & \bf $+$48 49 18.91 & \bf 21.50 & \bf 20.85 &  \bf 0.61 & \bf 0.427146 & \bf 0.000093 &  \bf 9.43/\nodata & \bf Y/G1  \\
 \bf 1340 & \bf 10 38 47.508 & \bf $+$48 50 19.14 & \bf 20.59 & \bf 20.12 &  \bf 0.57 & \bf 0.461146 & \bf 0.000180 &  \bf 7.29/\nodata & \bf N  \\
 \bf 1362 & \bf 10 38 47.760 & \bf $+$48 50 39.52 & \bf 19.26 & \bf 18.68 &  \bf 0.59 & \bf 0.471193 & \bf 0.000140 &  \bf 6.88/\nodata & \bf N   \\
 \bf 1393 & \bf 10 38 48.372 & \bf $+$48 47 46.46 & \bf 20.86 & \bf 20.27 &  \bf 0.61 & \bf 0.429957 & \bf 0.000123 &  \bf 8.83/\nodata & \bf Y/G1   \\
 \bf 1406 & \bf 10 38 48.552 & \bf $+$48 48 30.85 & \bf 22.21 & \bf 21.59 &  \bf 0.68 & \bf 0.835114 & \bf 0.000190 &\bf \nodata/~~~~~2 & \bf N   \\
 \bf 1438 & \bf 10 38 49.020 & \bf $+$48 50 47.62 & \bf 22.27 & \bf 21.86 &  \bf 0.47 & \bf 0.445005 & \bf 0.000163 &  \bf 5.90/\nodata & \bf N   \\
 \bf 1435 (B2) & \bf 10 38 49.469 & \bf $+$48 48 51.24 & \bf 21.89 & \bf 21.43 & \bf 0.56 & \bf 0.426500 & \bf 0.000600 &\bf \nodata        & \bf Y/G1  \\
 \bf 1454 & \bf 10 38 49.740 & \bf $+$48 48 18.65 & \bf 20.21 & \bf 19.82 &  \bf 0.37 & \bf 0.472110 & \bf 0.000193 &  \bf 8.09/\nodata & \bf N   \\
 \bf 1576 & \bf 10 38 51.432 & \bf $+$48 48 12.78 & \bf 21.57 & \bf 21.00 &  \bf 0.59 & \bf 0.433917 & \bf 0.000270 &  \bf 2.30/\nodata & \bf Y/G2   \\
 \bf 1605 & \bf 10 38 51.432 & \bf $+$48 48 48.74 & \bf 22.66 & \bf 22.01 &  \bf 0.61 & \bf 0.212771 & \bf 0.000113 &  \bf 6.34/\nodata & \bf N  \\
 \bf 1544 & \bf 10 38 51.972 & \bf $+$48 48 00.94 & \bf 20.94 & \bf 20.67 &  \bf 0.39 & \bf 0.434270 & \bf 0.000063 &\bf \nodata/~~~~~4 & \bf Y/G2  \\
 \bf 1696 & \bf 10 38 53.088 & \bf $+$48 48 25.81 & \bf 22.89 & \bf 22.37 &  \bf 0.33 & \bf 0.791988 & \bf 0.000140 &\bf \nodata/~~~~~4 & \bf N  \\
 \bf 1792 & \bf 10 38 54.168 & \bf $+$48 48 51.84 & \bf 22.27 & \bf 21.66 &  \bf 0.60 & \bf 0.432626 & \bf 0.000150 &  \bf 5.88/\nodata & \bf Y/G2  \\
  B3 &  10 38 24.822 & +48 49 56.384 &  \nodata &  \nodata &  \nodata &  0.43048 &  0.00016 &  \nodata &  Y/G2 \\
 2043 & 10 38 28.248 & $+$48 48 51.98 & 22.70 & 22.21 &  0.40 & 0.260293 & 0.000157 &\nodata/~~~~~2 & N  \\
 2049 & 10 38 28.320 & $+$48 48 56.23 & 21.83 & 21.12 &  0.70 & 0.470996 & 0.000103 &  9.13/\nodata & N  \\
 2080 & 10 38 29.436 & $+$48 50 28.32 & 22.57 & 21.68 &  0.69 & 0.890203 & 0.000077 &\nodata/~~~~~4 & N   \\
 2107 & 10 38 29.508 & $+$48 48 59.90 & 21.57 & 21.19 &  0.35 & 0.420291 & 0.000200 &\nodata/~~~~~6 & N   \\
 1356 & 10 38 31.272 & $+$48 49 48.32 & 21.51 & 20.95 &  0.59 & 0.430565 & 0.000123 &  6.64/\nodata & Y/G2  \\
 2184 & 10 38 31.308 & $+$48 50 06.68 & 19.18 & 18.86 &  0.40 & 0.208317 & 0.000267 &  7.29/\nodata & N  \\
  381 & 10 38 32.424 & $+$48 47 43.87 & 22.34 & 21.80 &  0.54 & 0.431942 & 0.000137 &  5.63/\nodata & Y/G2   \\
  450 & 10 38 33.396 & $+$48 47 45.42 & 22.81 & 22.34 &  0.49 & 0.419921 & 0.000203 &  2.84/\nodata & N   \\
  517 & 10 38 34.260 & $+$48 47 40.49 & 22.37 & 21.86 &  0.47 & 0.472610 & 0.000107 &\nodata/~~~~~6 & N   \\
  644 & 10 38 36.744 & $+$48 51 27.58 & 22.47 & 21.95 &  0.55 & 0.433904 & 0.000160 &  4.31/\nodata & Y/G2   \\
  778 & 10 38 38.292 & $+$48 51 35.71 & 20.24 & 19.76 &  0.49 & 0.360463 & 0.000060 &\nodata/~~~~~7 & N \\
  1096 & 10 38 44.268 & $+$48 51 19.51 & 20.19 & 19.69 &  0.61 & 0.431825 & 0.000140 &  6.30/\nodata & Y/G2  \\
   B4  & 10 38 44.357 & +48 52 13.602 & \nodata &  \nodata &  \nodata & 0.11716 & 0.00050 & \nodata  & N \\
  1112 (B5) & 10 38 44.990 & $+$48 47 15.43 & 17.47 & 17.01 & 0.48 & 0.176120 & 0.000500 & \nodata & N \\
  1181 & 10 38 45.420 & $+$48 46 59.34 & 22.84 & 22.08 &  0.34 & 0.493398 & 0.000050 &\nodata/~~~~~4 & N  \\
   B6 & 10 38 47.589 & +48 45 52.831 & \nodata & \nodata & \nodata & 0.43289 & 0.00034 & \nodata & Y/G2 \\
  1385 & 10 38 49.020 & $+$48 47 17.56 & 21.82 & 21.28 &  0.67 & 0.451399 & 0.000167 &  6.49/\nodata & N  \\
   B7  & 10 38 49.644 & +48 52 20.771 & \nodata &  \nodata &  \nodata & 0.44164 & 0.00006 & \nodata & N  \\
  1674 & 10 38 52.764 & $+$48 48 10.04 & 22.34 & 21.68 &  0.58 & 0.994848 & 0.000180 &\nodata/~~~~~2 & N  \\
  1681 & 10 38 52.764 & $+$48 47 33.43 & 21.23 & 20.59 &  0.62 & 0.433710 & 0.000083 & 11.96/\nodata & Y/G2  \\
  1711 & 10 38 53.484 & $+$48 50 07.44 & 21.60 & 21.06 &  0.52 & 0.530524 & 0.000267 &  4.68/\nodata & N  \\
  1787 & 10 38 54.420 & $+$48 50 40.09 & 21.29 & 20.63 &  0.68 & 0.434060 & 0.000077 & 12.58/\nodata & Y/G2  \\
  1782 & 10 38 54.708 & $+$48 50 13.63 & 22.69 & 21.90 &  0.72 & 0.891784 & 0.000127 &\nodata/~~~~~2 & N  \\
  1847 & 10 38 55.068 & $+$48 50 37.82 & 22.66 & 22.24 &  0.63 & 0.892748 & 0.000077 &\nodata/~~~~~4 & N  \\
   253 & 10 38 55.536 & $+$48 49 56.10 & 21.61 & 20.76 &  0.80 & 0.542742 & 0.000093 & 12.37/\nodata & N   \\
   228 & 10 38 55.716 & $+$48 49 49.40 & 20.89 & 20.40 &  0.42 & 0.239949 & 0.000017 &\nodata/~~~~~8 & N  \\
   208 & 10 38 55.932 & $+$48 47 18.71 & 19.85 & 19.39 &  0.49 & 0.190752 & 0.000294 &  2.75/\nodata & N  \\
   100 & 10 38 56.940 & $+$48 50 04.63 & 20.30 & 19.99 &  0.34 & 0.296108 & 0.000083 &\nodata/~~~~~8 & N  \\
\enddata
\label{tbl:tabgal}
\end{deluxetable*}

\begin{deluxetable*}{lccccccccl}
\tabletypesize{\scriptsize}
\tablecaption{Redshifts and magnitudes of the observed galaxies}
\tablenum{1}
\tablecolumns{10}
\tablewidth{0pc}
\tablehead{
\colhead{ID} &
\colhead{RA} &
\colhead{DEC} &
\colhead{\sloanr} &
\colhead{\sloani} &
\colhead{\sloanri} &
\colhead{$z$} &
\colhead{$\Delta (z)$} &
\colhead{R/\#em. lines} &
\colhead{Member?} \\
\colhead{(1)} &
\colhead{(2)} &
\colhead{(3)} &
\colhead{(4)} &
\colhead{(5)} &
\colhead{(6)} &
\colhead{(7)} &
\colhead{(8)} &
\colhead{(9)} &
\colhead{(10)}}
\startdata
   63 &  10 38 57.660 &  $+$48 47 33.32 &  21.10 &  20.43 &   0.75 &  0.416682 &  0.000130 &   4.56/\nodata &  N  \\
    1 & 10 38 57.948 & $+$48 50 13.99 & 21.42 & 20.70 &  0.78 & 0.502117 & 0.000287 &  7.91/\nodata & N  \\
  102 & 10 38 58.092 & $+$48 50 16.62 & 22.35 & 21.59 &  0.75 & 0.502978 & 0.000247 &  3.06/\nodata & N  \\
   72 & 10 38 58.488 & $+$48 46 56.06 & 20.46 & 19.79 &  0.73 & 0.435771 & 0.000103 &  9.05/\nodata & Y/G2  \\
    2 &  10 38 58.668 &  $+$48 47 06.58 &  20.89 &  20.22 &   0.66 &  0.435345 &  0.000223 &   5.02/\nodata &  Y/G2   \\
    B8 & 10 39 00.638 & +48 48 34.769 & \nodata &  \nodata &  \nodata & 0.20626 & 0.00050 & \it \nodata & N  \\
   \it B9 & \it 10 38 14.912 & \it +48 52 04.017 & \it \nodata & \it \nodata & \it \nodata & \it 0.42786 & \it 0.00013 & \it \nodata & \it Y/G1 \\
   \it B10 & \it 10 38 15.809 & \it +48 52 59.636 & \it \nodata & \it \nodata & \it \nodata & \it 0.43042 & \it 0.00013 & \it \nodata & \it Y/G2 \\
   \it B11 & \it 10 38 25.802 & \it +48 44 42.486 & \it \nodata & \it \nodata & \it \nodata & \it 0.08975 & \it 0.00050 & & \it N \\
   \it B12 & \it 10 38 37.863 & \it +48 53 25.550 & \it \nodata & \it \nodata & \it \nodata & \it 0.42949 & \it 0.00009 & \it \nodata & \it Y/G1 \\
   \it B13 & \it 10 38 42.254 & \it +48 53 14.399 & \it \nodata & \it \nodata & \it \nodata & 0.11697 & \it 0.00050 & \it \nodata & \it N \\
   \it B14 & \it 10 38 44.615 & \it +48 54 15.524 & \it \nodata & \it \nodata & \it \nodata & \it 0.42546 & \it 0.00022 & \it \nodata & \it Y/G1 \\
   \it B15 & \it 10 38 57.378 & \it +48 45 40.897 & \it \nodata & \it \nodata & \it \nodata & \it 0.37074 & \it 0.00010 & \it \nodata & \it N \\
   \it B16 & \it 10 39 02.732 & \it +48 46 35.045 & \it \nodata & \it \nodata & \it \nodata & \it 0.42704 & \it 0.00016 & \it \nodata & \it Y/G1 \\
   \it B17 & \it 10 39 05.797 & \it +48 53 24.039 & \it \nodata & \it \nodata & \it \nodata & \it 0.43186 & \it 0.00016 & \it \nodata & \it Y/G2 \\
   \it B18 & \it 10 39 05.940 & \it +48 46 45.029 & \it \nodata & \it \nodata & \it \nodata & \it 0.43386 & \it 0.00011 & \it \nodata & \it Y/G2 \\
   \it B19 & \it 10 39 07.866 & \it +48 48 09.185 & \it \nodata & \it \nodata & \it \nodata & \it 0.40277 & \it 0.00009 & \it \nodata &  \it N \\
   \it B20 & \it 10 39 09.609 & \it +48 49 59.268 & \it \nodata & \it \nodata & \it \nodata & \it 0.42759 & \it 0.00011 & \it \nodata & \it Y/G1 \\
   \it B21 & \it 10 39 10.594 & \it +48 50 23.643 & \it \nodata & \it \nodata & \it \nodata & \it 0.42996 & \it 0.00012 & \it \nodata & \it Y/G1 \\
   \it B22 & \it 10 39 14.033 & \it +48 51 28.229 & \it \nodata & \it \nodata & \it \nodata & \it 0.44482 & \it 0.00014 & \it \nodata &  \it N \\
   \it B23 & \it 10 39 18.069 & \it +48 48 05.024 & \it \nodata & \it \nodata & \it \nodata & \it 0.42894 & \it 0.00009 & \it \nodata & \it Y/G1 \\
\enddata
\tablecomments{Galaxies within 0.5 $r_{200}$, 0.5 -- 1.0 $r_{200}$, and 1.0 -- 1.5 $r_{200}$ of the X-ray centroid of the Cheshire Cat shown
in bold, regular font, and italics, respectively. Column (1): source Extractor Galaxy ID, with B1-B23 from \citet{bayliss14};
Columns (2), (3): right Ascension and Declination (J2000.0).
The units of Right Ascension are hours, minutes and seconds, and the units of Declination are degrees, arcminutes
and arcseconds; Columns (4) and (5): total magnitudes in \sloanr~and \sloani; Column (6): Sloan \sloanr - \sloani color measured inside a
fixed aperture of 1\farcs2; Columns (7) and (8): redshift and the associated error, corrected to the heliocentric system;
Column (9): the $R$ value \citep{tonry79} and/or the emission lines used to calculate the redshift; Column (10): membership
flag: Y/G1 - member galaxy of G1, Y/G2 - member galaxy of G2, N - background/foreground galaxy.}
\end{deluxetable*}

\begin{figure}
\hspace{-10mm}
\vspace{-3mm}
 \includegraphics[height=9.4cm,width=9.4cm]{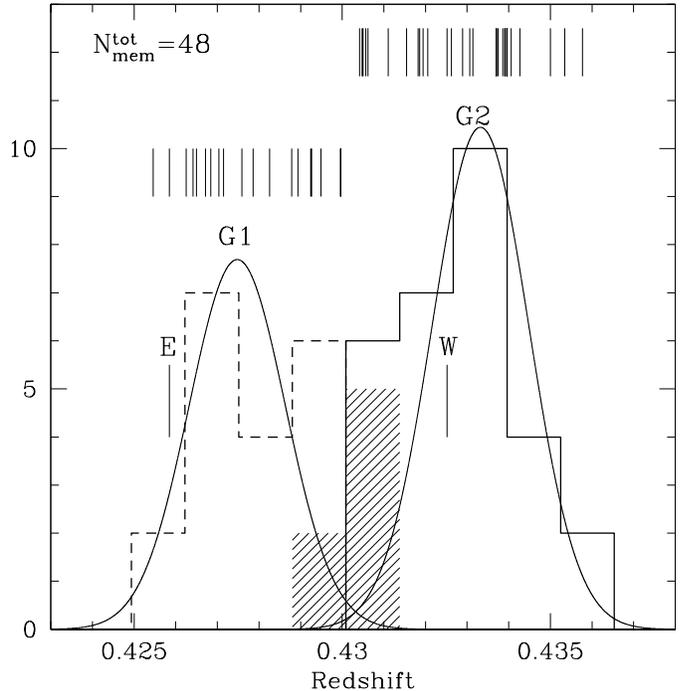}
 \caption{Histogram of the redshift distribution of 48 spectroscopically confirmed galaxies belonging to the Cheshire Cat.
Galaxies have been separated into the eastern eye group (G1, 19 galaxies; dashed line) and western eye group
(G2, 29 galaxies; solid line). The shaded histogram represents the seven galaxies with posterior probabilities of  group
memberships between 58\%~and 62\% (2 galaxies in G1 and 5 galaxies in G2; see text). The smooth curves represent the best fit to
a bimodal velocity distribution. The individual galaxy redshifts are shown as vertical lines at the top of the Figure.
``E" and ``W" represent the redshifts of the eastern and western eye galaxies, respectively.}
\label{fig:histo}
\end{figure}

Combining our group members with those in the \citet{bayliss14} sample, there are 48 galaxies in the redshift interval 
between $0.424 < z < 0.437$ that we identify as being part of the Cheshire Cat system.  Thirty-six of the
galaxies are located inside the GMOS field of view, which corresponds to the virial radius of the group of
1.3 Mpc (see below), and 12 are located between 1.0--1.5 virial radii (1.3 -- 2.0 Mpc). We use only the galaxies
inside the virial radius to calculate the virial mass and group richness values, while we use all 48 galaxies to
derive the redshift distribution.
The redshift distribution of these galaxies is shown in Figure~\ref{fig:histo}. Two prominent peaks are visible in the redshift distribution,
suggesting that there could be more than one structure along the line of sight of the 
group. In order to investigate its structure, we used the KMM test 
\citep{ashman94}, which is appropriate to detect the presence of two or more components in an 
observational data set.  We have considered whether the data is consistent with a single component or not.  
The results of applying the test in the homoscedastic mode (common covariance) yields strong evidence 
that the redshift distribution of member galaxies is at least bimodal, rejecting a single Gaussian model 
at a confidence level of 98.10\% ($P$-value of 0.019). The $P$-value is another way to express the statistical significance 
of the test, and is the probability that a likelihood test statistic would be at least as large as the observed value 
if the null hypothesis (one component in this case) were true. Assuming two components, the procedure assigns
a mean value of $z \sim 0.428$ and $z \sim 0.433$ with 19 (41\%) and 29 (59\%) galaxies for the structures,
respectively (groups we designate G1 and G2 in Figure~\ref{fig:histo}). We used the posterior probability of group membership 
given by KMM to assign the galaxies to each group. Seventeen out of 19 galaxies in group G1 and 24 out of 29 galaxies
in group G2 have a probability $>$ 80\% to be a members of these groups. Two of the galaxies in group G1
have a probability $\sim$63\%~to be members of this group. In the case of group G2, there are five galaxies
with a probability between 58\%\ and 63\% to be members of this group. It can not be ruled out the possibility
that these galaxies are part of a smaller group located between G1 and G2 groups or are part of a filament 
connecting the two structures. The membership flag and galaxy group assignments are given in the last 
column of Table~\ref{tbl:tabgal}. The seven galaxies with lower probability to be members of groups G1 and G2 
are represented by the shaded histogram in Figure~\ref{fig:histo}.

We have used the bi-weight estimator for location and scale \citep{beers90} to calculate the average redshift and 
the line of sight velocity dispersion of the Cheshire Cat group as as whole.  We used an iterative procedure by 
calculating the location and scale using the Robust Statistics (ROSTAT) program \citep{beers90} and applying a 3$\sigma$ clipping algorithm to the 
results. We repeated this procedure until the velocity dispersion converged to a constant value. We calculate an average redshift for
the group of $0.430877\pm0.000307$ and a line of sight velocity dispersion of $\sigma_{los} = 659 \pm 69$
km s$^{-1}$. Given this velocity dispersion
 and using the prescription of \citet{heisler85}, the virial mass
is $(1.5 \pm 0.2) \times 10^{14}$ \Msol, and the virial radius is $1.3 \pm 0.1$ Mpc, with uncertainties
at the 68\% confidence intervals. We do the same thing for G1 and G2 separately, and
find average redshifts of 0.427751$\,\pm\,0.000250$ and 0.432867$\,\pm\,0.000213$, respectively, with
with line of sight velocity dispersions of $\sigma_{los} = 318 \pm 51$ km s$^{-1}$ and $\sigma_{los} = 337 \pm 42$ km s$^{-1}$,
respectively (see Figure~\ref{fig:histo}). The two groups are separated by $\sim$1350 km s$^{-1}$
in the group rest frame. The virial masses are $(0.33 \pm 0.07) \times 10^{14}$ \Msol~and
$(0.39 \pm 0.06) \times 10^{14}$ \Msol, while the virial radii are $0.64 \pm 0.07$ Mpc and $0.69 \pm 0.08$ Mpc for groups G1 and G2, respectively.
The velocities of the eastern and western eye galaxies are shown with an "E" and ``W," respectively. 
Curiously, neither eye is at the center of its respective group in velocity space. The eastern eye is offset to a 
lower velocity by $\approx$570 km s$^{-1}$, while the western eye is offset by only $\approx$100 km s$^{-1}$.

\begin{figure}
\hspace{-7mm}
\vspace{-3mm}
 \includegraphics[height=9.2cm,width=9.2cm]{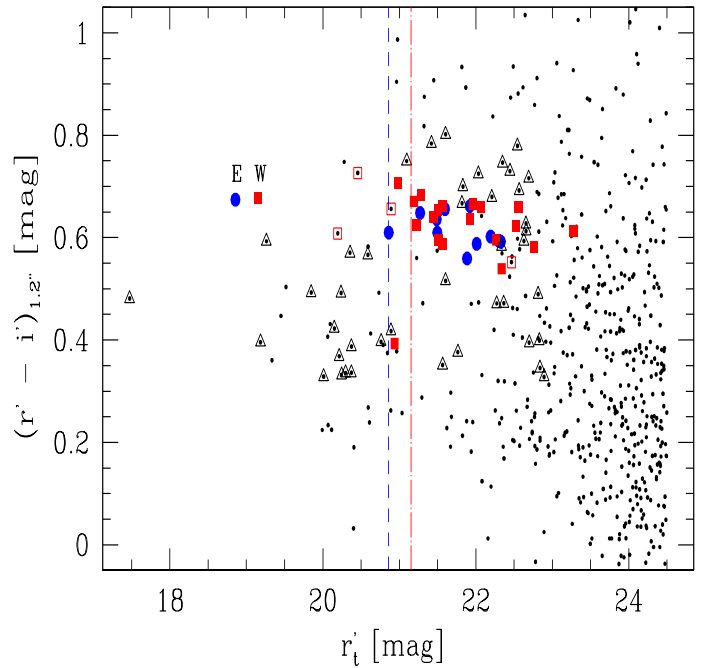}
 \caption{Color-magnitude diagram of all galaxies detected in the GMOS images with \sloanr brighter than 24.5 mag
(595 galaxies, black dots). Blue circles and red squares are the spectroscopically confirmed members of the
G1 and G2 groups respectively, with filled and open symbols representing which of these galaxies lie inside and outside of
0.5 $r_{200}$ of the group. Red short dash and blue long dash vertical lines represent a two magnitude gap from the E and W eye
galaxies, respectively. The black triangles represent spectroscopically confirmed background and foreground galaxies, while
black dots represent objects without spectroscopic redshifts.
The two central galaxies of the groups, SDSS J103843.58+484917.7 and SDSS J103842.68+484920.2 (the  ``eyes''
of the Cheshire Cat) are represented by a letter (E and W) in the CMD.} \label{fig:cmd}
\end{figure}

Figure~\ref{fig:cmd} shows the color magnitude diagram (CMD) of the 595 galaxies brighter than \sloanr = 24.5
mag (black dots). The blue circles and the red squares represent the spectroscopically confirmed
members of the G1 and G2 groups, respectively. Filled symbols represent galaxies within 0.5 $r_{200}$ of the combined system, while open symbols represent
galaxies outside 0.5 $r_{200}$. In the Figure we can see that the red sequence for galaxies members 
of the group show a moderate scatter in color. We also can see that galaxies members of the G1 group are slightly bluer 
than the galaxies members of the G2 group. There is a two magnitude gap in \sloanr~ between the G1 central galaxy (galaxy `E'; ID 972) and
the next brightest confirmed G1 galaxy (ID 1393) within 0.5 $r_{200}$ (0.32 Mpc) of the G1 center. We note that there are two galaxies within this radius that are within
two magnitudes in \sloanr~ of galaxy `E' for which we do not have a spectroscopic redshift. However, one is blue in color (\sloanr -- \sloani $<$ 0.5).
Of the 24 galaxies with spectroscopic redshifts that have colors this blue, only one galaxy is a member of either G1 or G2.
Thus, it is statistically unlikely that this galaxy actually belongs to the Cheshire Cat. The other galaxy is red, but has an SDSS-determined photometric
measurement that is $>$3$\sigma$ larger than the redshift of the Cheshire Cat. So G1 was likely a fossil group before its collision with G2, although further optical
spectroscopy will be needed for these galaxies. Similar arguments lead to a gap of 1.83
magnitude for the G2 central galaxy (``W"; ID 1092) and the second rank galaxy within 0.5 $r_{200}$ (0.35 Mpc) of the G2 center (ID 983), indicating G2 might have been a near-fossil group before collision with G1.
The two magnitude gap between the first- and second-ranked galaxies in a group ($\Delta m_{12}$) is not the only criterion that has been used to identify
fossil groups. \citet{dariush10} suggest a 2.5 magnitude gap between the first- and fourth-ranked galaxies ($\Delta m_{14}> 2.5$) as an alternative means of identifying
early-formed systems. With this definition, G1 was a fossil group unless both galaxies without spectroscopic redshifts mentioned above are G1 members. G2 would once
again be a near-fossil group with $\Delta m_{14} = 1.83$ at worst, or $\Delta m_{14} = 2.36$ if the galaxies without spectroscopic redshifts are ignored\footnote[9]{As will be shown in
\S~\ref{sec:hst}, the BGG of G2 actually has a double nucleus with a separation of only 0\farcs37 (2.1 kpc). Since this double nucleus is not apparent in our
{\it Gemini} GMOS imaging, the quoted magnitude of this galaxy combines the luminosities of both nuclei. Our determination of $\Delta m_{12}$ and $\Delta m_{14}$ therefore do not
consider the two nuclei to be two separate galaxies.}
This will become
relevant in \S~\ref{ssec:future} when we discuss the past and future of the Cheshire Cat system once this suspected merger is complete.
% We notice also that the central galaxy of the G1 group SDSS J103843.58+484917.7 
%(E)  at least $\sim$2 mag brighter than the second ranked galaxy in the group. In the case of G2 group, its central galaxy 
%SDSS J103842.68+484920.2 (W) is only $\sim$ 1.3 mag fainter than the second ranked galaxy, suggesting that this
%structure is not a Fossil Group.}

%\begin{figure}
%\hspace{-7mm}
%\vspace{-3mm}
% \includegraphics[height=9.2cm,width=9.2cm]{newfigcmd.eps}
% \caption{Color-magnitude diagram of all galaxies detected in the GMOS images with \sloanr~brighter than 24.5 mag
%(595 galaxies, black dots). Blue circles and red squares are the spectroscopically-confirmed members of the
%G1 and G2 groups respectively, with filled and open symbols representing which of these galaxies lie inside and outside of
%0.5 $r_{200}$ of the group. Red short dash and blue long dash vertical lines represent a two magnitude gap from the E and W eye
%galaxies, respectively. The black triangles represent spectroscopically-confirmed background and foreground galaxies, while
%black dots represent objects without spectroscopic redshifts.
%The two central galaxies of the groups, SDSS J103843.58+484917.7 and SDSS J103842.68+484920.2 (the  ``eyes''
%of the Cheshire Cat) are represented by a letter (E and W) in the CMD.} \label{fig:cmd}
%\end{figure}

We can use our CMD to estimate $N_{200}$, the number of red ridge galaxies brighter than 0.4 $L^{\star}$ that are within
a radius of $r_{200}$. For the Cheshire Cat, 0.4 $L^{\star}$ corresponds to an apparent \sloani~magnitude of
20.84 \citep{wiesner12}. Using Sloan and WIYN data, \citet{wiesner12} estimate that $N_{200} = 25$ for the
Cheshire Cat based on broad band optical colors.
%As we show in \S~\ref{sec:mass}, a group with $N_{200} = 25$ would be expected to have $r_{200}$ of 1.3 Mpc.
Our {\it Gemini} GMOS imaging covers most of a 1.3 Mpc radius circle around
the group, except for small regions to the extreme north and south of the 1.3 Mpc circle. Thus, we are able to
image nearly every red ridge galaxy within $r_{200}$.
From our spectroscopic data, we confirm 20 red ridge galaxies in the Cheshire Cat within 1.3 Mpc whose \sloani~magnitude in the
Sloan Catalog ({\it not} in our catalog, since $N_{200}$ was calibrated for SDSS magnitudes)
is brighter than 20.84.
From Figure~\ref{fig:cmd}, there are a few galaxies that lie in the domain of the
brighter red ridge group members that do not have redshift measurements.
Given the purity of the red ridge sample in this area, it is likely
that 2--3 of these galaxies belong to the Cheshire Cat and should be counted in the $N_{200}$ determination. An additional
bright red ridge galaxy might be expected statistically in the small portion of the 1.3 Mpc circle that is not covered by our
observations to the north and
to the south of the group. Thus, there are most likely 23--24 red ridge galaxies brighter than \sloani =  20.84 within 1.3 Mpc, consistent with
\cite{wiesner12}. Separating the bright red ridge galaxies into either the G1
or G2 groups within their respective $r_{200}$ determined above, we estimate $N_{200,G1}=7$ and $N_{200,G2}=13$.
%From \S~\ref{sec:mass}, this implies $r_{200,G1}=0.64$ Mpc and $r_{200,G2}=0.86$ Mpc.
%While the smaller number of galaxies per group implies a smaller $r_{200}$ value for both groups, no further iteration is necessary to determine
%$N_{200,G1}$ and $N_{200,G2}$ since no additional red ridge galaxies are lost between 1 Mpc and the new $r_{200}$ radii.

\section{{\it Chandra} X-ray Data Analysis}
\label{sec:xray}

Previously unobserved in X-rays, {\it Chandra} ACIS-S observations (ObsID 11756 and 12098)
for a combined 70.2 ks exposure were obtained on 2010 January 24 and 2010 February 1 for the Cheshire Cat.
The data sets were processed beginning with the level 1 event file in a uniform manner following the
{\it Chandra} data reduction threads employing {\tt CIAO}~4.5 coupled with
{\tt CALDB}~4.5.7. Events with grades of 0, 2, 3, 4 and 6 were selected for all subsequent
processing and analysis. The observation specific bad pixel files were applied
from the standard calibration library included with {\tt CALDB}~4.5.7.

\begin{figure*}
\hspace{20mm}
 \includegraphics[height=15.5cm,width=14.5cm]{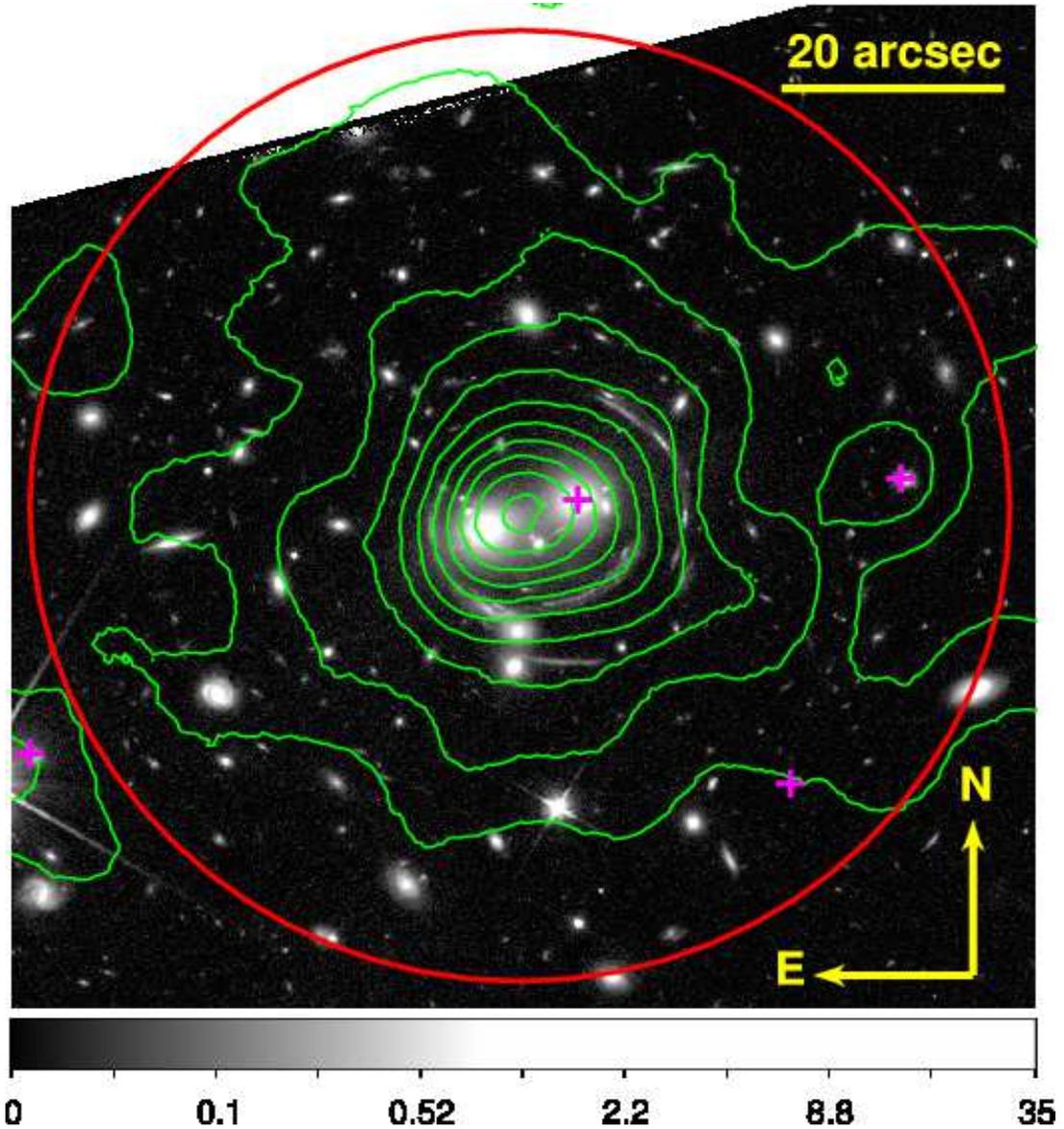}
\vspace{6mm}
 \caption{Adaptively smoothed 0.3--7.0 keV {\it Chandra} ACIS-S contours (green) of the Cheshire Cat plotted over the F110W {\it HST} image showing
extended, diffuse X-ray emission. Magenta crosses represent the locations of X-ray point sources. The red circle represents 0.2 $r_{200}$.}
% While not evident in the image due to the smoothing necessary to enhance the diffuse X-ray emission, The red circle represents $0.2 r_{200}$.
%the western eye is coincident with an X-ray point source that is likely an AGN.}}
\label{fig:chandra}
\end{figure*}

\begin{figure*}
\hspace{20mm}
 \includegraphics[height=14.5cm,width=15.5cm]{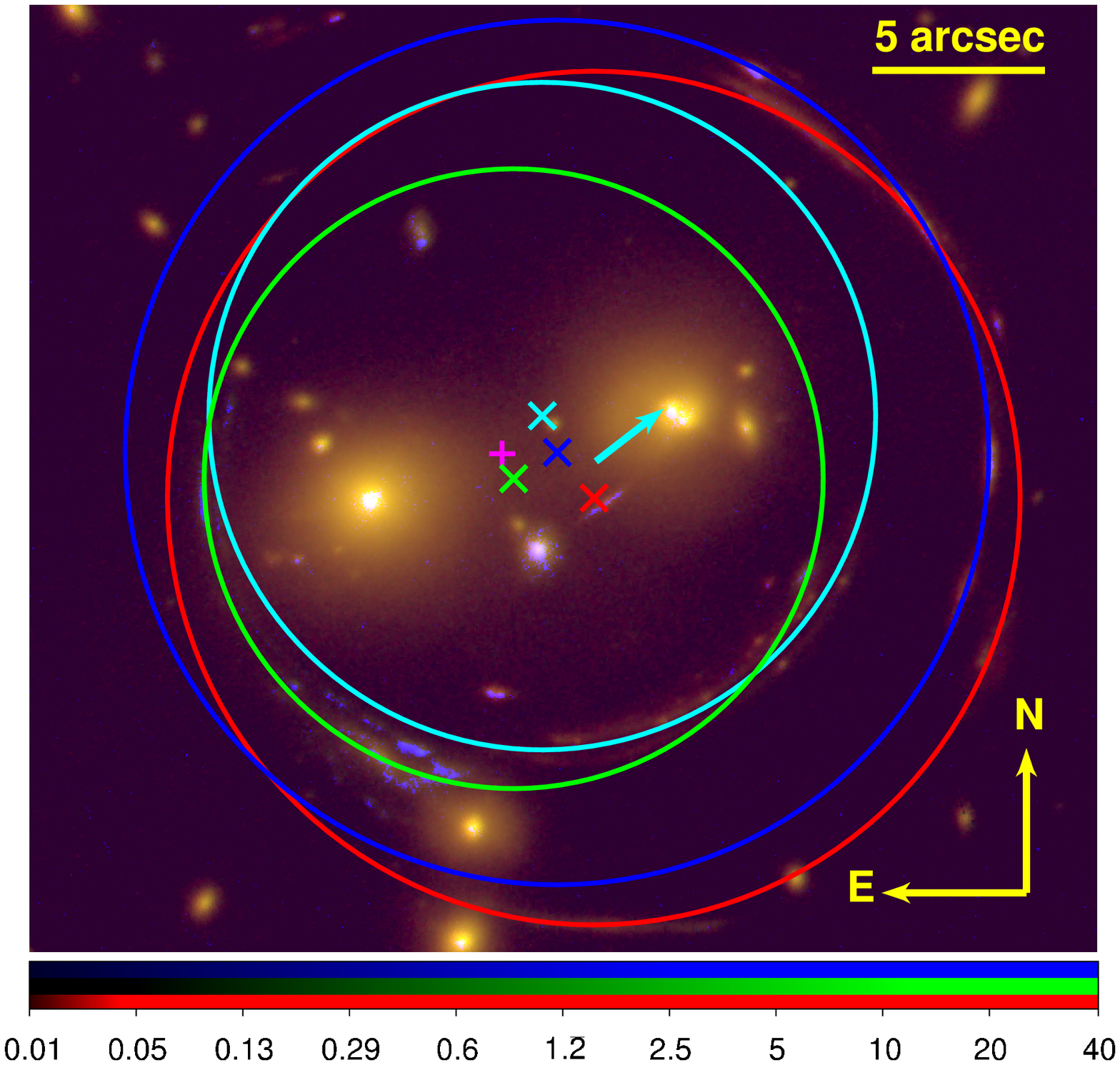}
\vspace{6mm}
 \caption{{\it HST} image of the Cheshire Cat group with F390W, F110W and F160W filters as blue, green and red in log scale.
% Units are counts per second.
Likely arc systems are indicated using green, blue, red and cyan circles.  The center of each circle is indicated with an ``x"
of the same color. The X-ray centroid, for comparison, is indicated with a magenta cross. Also note the off-center double nucleus of
the western eye. A bright X-ray point source is associated with the northeast (WA) nucleus (cyan arrow).}
\label{fig:circlefit}
\end{figure*}

There were no times of excessively high background, so the entire 70.2 ks observation time was utilized. The two
individual observations were merged using the {\tt CIAO} tool {\tt reproject\_obs } for the spatial analysis. X-ray
point sources were detected using {\tt wavdetect} and verified by eye. For spatial work, the surface brightness was
extracted using {\tt SHERPA} in annular rings centered on the peak of X-ray emission. Background was selected from a
hot gas-free and point source-subtracted annular region centered on the peak of emission with an extent of
120$^{\prime\prime}-150^{\prime\prime}$.
For spectral work, we used
{\tt specextract} to extract spectra and generate {\tt RMF} and {\tt ARF} files for each observation separately, and then combine
them into one spectrum by appropriately weighting the two {\tt RMF}s and {\tt ARF}s. Spectra were fit within XSPECv12.8
using an {\tt apec} thermal model with Galactic absorption \citep[using {\tt tbabs} and a value of $N_H = 1.35 \times 10^{20}$ cm$^{-2}$;][]{dickey90}
using $\chi^2$ statistics with energy channels grouped such that each energy bin contained
at least 25 photons. The abundance table from \citet{grevesse98} was adopted.
Background was selected from the same region used in the spatial analysis.
Spectra were fit in the 0.5--7.0 keV energy range.

\begin{figure}
\hspace{-10mm}
\vspace{-3mm}
 \includegraphics[height=9.5cm,width=9.5cm]{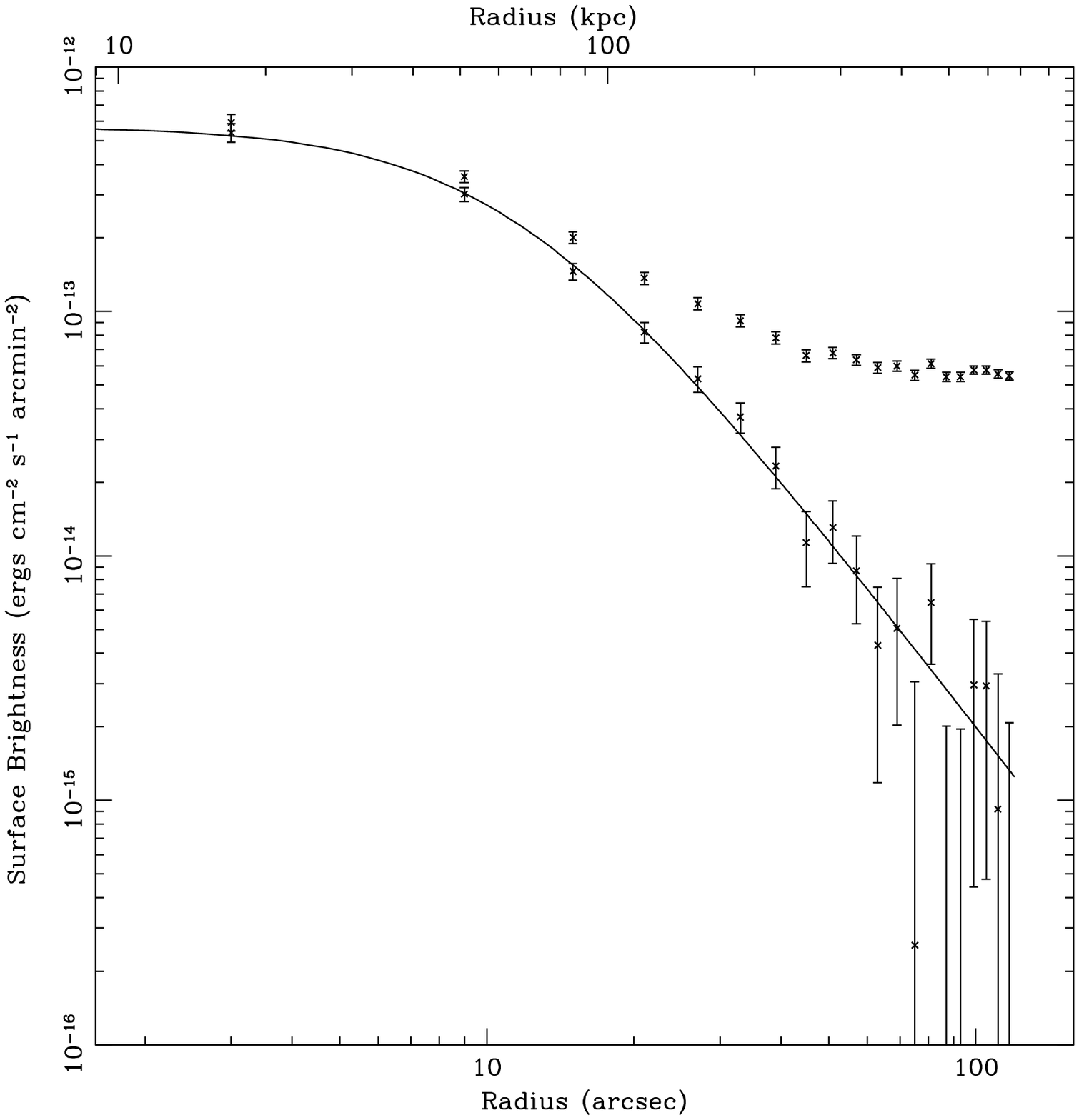}
 \caption{Background-subtracted and non-background-subtracted 0.5--7.0 keV X-ray surface brightness profile of the Cheshire Cat group
with 1${\sigma}$ uncertainties. The
best-fit $\beta$-profile is also shown with parameters $\beta = 0.60 \pm 0.03$ and core radius
$r_c = 64.7^{+8.5}_{-7.4}$ kpc.}
\label{fig:xsurf}
\end{figure}

The adaptively smoothed 0.3--7.0 keV X-ray contours are shown plotted over the {\it HST} image in Figure~\ref{fig:chandra}. Diffuse X-ray emission is
detected coincident with the group with a centroid of $(\alpha,\delta)=(10^{h}38^{m}43^{s}.196$, $+48\degr49\arcmin19\arcsec.09)$. By computing the
X-ray centroid on  a variety of spatial scales from 5$^{\prime\prime}$ to 60$^{\prime\prime}$ and noting the spread in center values, we estimate the
uncertainty of the X-ray centroid to be 2$^{\prime\prime}$. The centroid is not
located at the position of either eye, but approximately half-way between the eyes (see Figure~\ref{fig:circlefit}). A bright
X-ray point source is coincident with the western eye with a luminosity of $L_X = 1 \times 10^{43}$ erg s$^{-1}$ (0.3--10 keV) assuming it
is at the same redshift as the eye.
The background-subtracted X-ray surface brightness profile is shown in Figure~\ref{fig:xsurf}, with emission
detected reliably out to a radius of 120$^{\prime\prime}$ (570 kpc) with a total of about 1160 net counts. The profile was
fit within {\tt SHERPA} with a standard $\beta$-model with best-fit parameters $\beta = 0.60 \pm 0.03$ and a core-radius
$r_c = 64.7^{+8.5}_{-7.4}$ kpc (1$\sigma$ uncertainties), both typical values for groups of galaxies.

\begin{figure}
\hspace{-10mm}
\vspace{-3mm}
 \includegraphics[height=9.5cm,width=9.5cm]{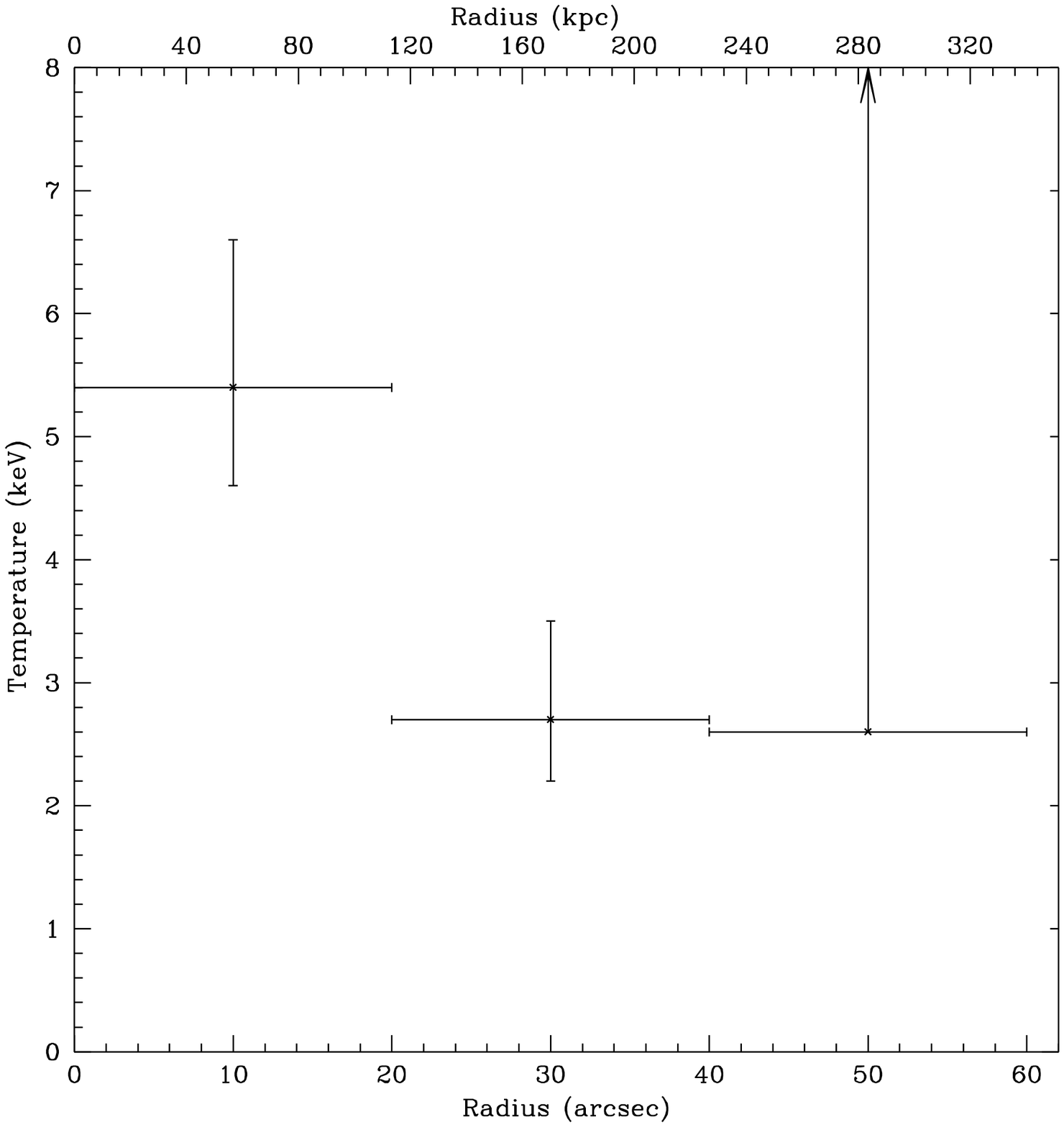}
\caption{Temperature profile of the hot gas in the Cheshire Cat with 1$\sigma$ confidence uncertainties.}
\label{fig:xtemp}
\end{figure}

The global temperature within a radius of 60$^{\prime\prime}$ (where with signal-to-noise begins to decline
significantly) is $4.3^{+0.9}_{-0.7}$ keV (1$\sigma$ uncertainties). No useful constraints could be placed on the metallicity
of the gas, so it was set to 40\% of the solar value for all subsequent work. The inner 60$^{\prime\prime}$ was broken
into three equally spaced annular regions to derive a temperature profile, shown in
Figure~\ref{fig:xtemp}. The central 20$^{\prime\prime}$ shows a strong peak ($5.4^{+1.2}_{-0.8}$ keV) before dropping
below 3 keV at larger radii. The total 0.3--10 keV (0.1--2.4 keV) luminosity within $120^{\prime\prime}$ is $1.1 (0.72) \times 10^{44}$ erg s$^{-1}$.
The central cooling time can be estimated from \citet{voigt04} given as $t_{cool} = 20 (n_e/10^{-3}~{\rm cm}^{-3})^{-1} (T/10^7~{\rm K})^{1/2}~{\rm Gyr}$.
Given a central temperature of 5.4 keV and a derived central density of $1.5 \times 10^{-2}$ cm$^{-3}$, the central cooling time is $\sim$3 Gyr.

\section{{\it HST} Imaging Data Analysis}
\label{sec:hst}

In order to study the eye galaxies and gravitational lenses of the Cheshire Cat in more detail, we examined archival observations taken
with {\it HST} WFC3 using the F390W, F110W and F160W filters (program ID 13003).  {\it HST} observed the
target for 2400s with F390W, 1100s with F110W, and 1200s with F160W on 2013 March 16.  The pivot wavelengths of these filters respectively
correspond to 2741, 8060, and 10740\;\AA\ in the rest frame.  We initially screened for cosmic rays using
{\tt lacosmic} \citep{lacosmic}. We aligned and co-added the filter sub-exposures using the {\tt astrodrizzle}
package \citep{drizzlepac}, with a scale of 0\farcs0588 per pixel and a pixel drop fraction of 0.6.

In order to estimate the center-of-mass for the system for comparison with results from the X-ray data, we examined the arc system of the Cheshire Cat.
\citet{belokurov09} and \citet{bayliss11} identified four major ring systems associated with background galaxies
at redshifts 0.80, 0.97, 2.20 and 2.78. The arc system displayed in {\it HST} imaging, however, is considerably more complex
than indicated in either previous study, with multiple overlapping arcs and knots where previously only single knots were assumed.
To first order, we assume that the center of mass for the lens is at the center of a circle which fits the surrounding system of arcs
and knots, as for a simple ``Einstein ring."  We then fit circles by eye which correspond approximately to a given arc system.
We begin with the assumption that \citet{belokurov09} and \citet{bayliss11} have recovered the correct arc systems, and then
modify the circles as needed, associating knots by color.

Each of these circles is centered within $\sim$1\farcs5 of
$(\alpha,\delta)=(10^{h}38^{m}43^{s}.079$, $+48\degr49\arcmin18\arcsec.39)$. The X-ray centroid (\S~\ref{sec:xray}) falls very
near the lens arc centroids (Figure~\ref{fig:circlefit}).
This result is consistent with the bulk of lensing matter being located at approximately the midpoint between the eastern and western eyes.
If, alternatively, we assumed that either eye is the center of a lens, we would expect to see additional knots at other points around the
eye, but we find no evidence for such counterpart knots.  New and deeper integral field unit spectroscopy could confirm the configuration
of these lensing systems, which in turn would allow a more comprehensive mass distribution reconstruction. Such spectroscopy might be
possible from the ground with optimal seeing, given a typical knot thickness of $\sim$$0.7\arcsec$. We note that a weak lensing study of this
system by \citet{oguri12} found an offset between the center of the inferred mass distribution and the BGG of the group, presumably the eastern eye.

Inspection of the western eye reveals it to be complex, with two distinct nuclei separated by 0\farcs37 (2.1 kpc)
embedded off-center in an extended optical halo (Figure~\ref{fig:circlefit}). We label the nucleus to the northeast WA and the nucleus to
the southwest WB.
Careful alignment of the {\it Chandra} and {\it HST} images indicates that the bright X-ray point source mentioned in
(\S~\ref{sec:xray}) is associated with the nucleus WA. This X-ray source is coincident with a 5.96 mJy VLA FIRST source.
WA also appears significantly brighter in the {\it HST} F390W band
image than WB, while the nuclei are of nearly equal brightness in the two long wavelength bands. Taken together this
indicates that WA harbors an AGN.

\section{Discussion}
\label{sec:discuss}

\subsection{The Mass and Size of the Cheshire Cat Group}
\label{sec:mass}

Our combined optical and X-ray study of the Cheshire Cat system gives us multiple means of determining the mass of the
group at various radii as well as characteristic size estimates of the group.
Beginning with the galaxy richness parameter $N_{200}=23$, $N_{200,G1}=7$, and $N_{200,G2}=13$,
that we determined in \S~\ref{sec:gemini}, we can use
scaling relations from the Northern Sky Optical Cluster Survey \citep[NoSOCS;][]{lopes09} survey to find the
predicted $r_{200}$ and $M_{200}$, and compare those values
to scaling relations involving measured X-ray quantities. \citet{lopes09} find

\begin{equation}
\ln (r_{200}) = 0.05 + 0.39 \ln (\frac{N_{200}}{25}) h_{70}^{-1}~{\rm Mpc}
\end{equation}
which gives $r_{200} = 1.02$ Mpc, $r_{200,G1} = 0.64$ Mpc, and $r_{200,G2} = 0.81$ Mpc. The \citet{lopes09}
relation for mass $M_{200}$ is

\begin{equation}
\ln (M_{200}) = 0.21 + 0.83 \ln (\frac{N_{200}}{25}) \times 10^{14} h_{70}^{-1}~{\rm M}_{\odot}
\end{equation}
to give masses of $M_{200}= 1.2 \times 10^{14}$ \Msol, $M_{200,G1}= 4.3 \times 10^{13}$ \Msol, and
$M_{200,G2}= 7.2 \times 10^{13}$ \Msol. Alternatively, the $M_{200}$--$N_{200}$ relation for maxBCG clusters
with a mean redshift of $z=0.25$ from \citet{johnston07}

\begin{equation}
M_{200} = 1.26~(\frac{N_{200}}{20})^{1.28} \times 10^{14} h_{70}^{-1}~{\rm M}_{\odot},
\end{equation}
gives masses of $M_{200}= 1.5 \times 10^{14}$ \Msol, $M_{200,G1}= 3.3 \times 10^{13}$ \Msol, and
$M_{200,G2}= 7.3 \times 10^{13}$ \Msol. The mass for G1 is consistent with the mass determined from
the velocity dispersion (\S~\ref{sec:gemini}), while the mass for G2 is about a factor of two larger than
the mass determined from the velocity dispersion.

The galaxy richness parameter can also predict the X-ray properties of the Cheshire Cat group. Once again from 
\citet{lopes09}:

\begin{equation}
\ln (L_X) = -1.85 + 1.59 \ln (\frac{N_{200}}{25}) \times 10^{44}h_{70}^{-1}~{\rm erg~s^{-1}},
\end{equation}
where $L_X$ is measured in the {\it ROSAT}~0.1--2.4 keV band, and

\begin{equation}
\ln (T_X) = 1.05 + 0.56 \ln (\frac{N_{200}}{25}) h_{70}^{-1}~{\rm keV},
\end{equation}
yielding estimates of $L_X = 1.4 \times 10^{43}$ erg s$^{-1}$ and $T_X$ = 2.7 keV
(if applied to each sub-group separately, we obtain $L_{X,G1} = 2.1 \times 10^{42}$ erg s$^{-1}$,
{$L_{X,G2} = 5.6 \times 10^{42}$ erg s$^{-1}$}, $T_{X,G1}$ = 1.4 keV, and $T_{X,G2}$ = 2.0 keV).
Compared to the measured
values of $L_X (0.1-2.4~{\rm keV}) = 7.2 \times 10^{43}$ erg s$^{-1}$ and $T_X$ = 4.3 keV, the Cheshire Cat is hotter and much
more luminous than predicted. We believe the elevated X-ray luminosity and temperature is the result of
shock heating from a merger of the two groups, and we will return to this issue in \S~\ref{ssec:merger}.
%Indeed, the predicted temperature for the  is consistent with the measured X-ray temperature of 2.7$^{+1.5}_{-0.7}$ keV in our second
%annular bin outside the core of the group which may be less affected by shock heating due to the collision,
%and might therefore more accurately represent the true viral temperature of the group.

We can use the best-fit hot gas temperature along with the best-fit $r_c$ and $\beta$ values from the X-ray surface brightness
distribution to estimate $r_{200}$ and total gravitational mass via \citet{helsdon03} and \citet{arnaud99}:

\begin{equation}
r_{200} = 0.81 (\frac{kT}{{\rm keV}})^{1/2} h_{70}^{-1} E(z)^{-1}~{\rm Mpc},
\end{equation}

where $E(z) = [\Omega_m(1+z)^3 + \Omega_{\Lambda}]^{1/2}$ and

\begin{equation}
M(r=r_{200}) = 8.08 \times 10^{14} \beta \frac{kT}{\rm 10 keV} \frac{r_{200}}{\rm Mpc} \frac{(r_{200}/r_c)^2}{1 + (r_{200}/r_c)^2} h_{70}^{-1}~{\rm M}_{\odot},
\end{equation}
respectively.
We calculate each quantity using the measured global temperature of 4.3 keV, and the temperature outside the central, hot
core beyond 110 kpc of 2.7 keV which might more accurately represent the unshocked virial temperature of the group. For $kT = 4.3$ keV,
we obtain $r_{200} = 1.36$ Mpc and $M_{200} = 2.8 \times 10^{14}$ \Msol, while for $kT = 2.7$ keV, we obtain
$r_{200} = 1.08$ Mpc and $M_{200} = 1.3 \times 10^{14}$ \Msol. Using a lower temperature of $kT = 2.7$ keV gives an estimate of
$M_{200}$ more in line with the estimate derived from the galaxy richness and our estimate from the galaxy velocities,
lending further credence to the notion that the hot
$>5$ keV gas in the center and the high X-ray luminosity result from shock heating.

We must compare the properties of the Cheshire Cat to cluster samples in the literature rather than groups,
since the global temperature of 4.3 keV exceeds that of
the hottest groups in group samples (and hotter than all 11 fossil groups in the \citet{miller12} sample).
With a bolometric X-ray luminosity of $L_{X,bol} = 1.1 \times 10^{44}$ erg s$^{-1}$, the Cheshire Cat lies below the best-fit
$L_{X,bol} - kT$ relation of \citet{vikhlinin09} and \citet{wu99},
indicating that it is hotter than expected for its X-ray luminosity (or alternatively stated, X-ray underluminous for its temperature).
This is not unexpected given that the system has likely undergone a recent collision that has shock heated the gas to higher temperature.
The Cheshire Cat lies slightly above the $L_X - M_{200}$ relation of \citet{rykoff08} and consistent with the fossil group sample of
\citep{miller12}). The summed \sloanr~magnitudes of the confirmed galaxies is $6 \times 10^{11}$ L$_{\odot}$ putting the
Cheshire Cat somewhat above the $L_{opt} - L_{X,bol}$ relation of \cite{girardi14} for non-fossil groups, but within the scatter for fossil groups.
Given that the Cheshire Cat is believed to be a merger, it is useful to compare its properties to the scaling relations of merging groups and clusters.
\cite{kettula14} recently determined scaling relations for a sample of clusters from weak lensing and X-ray analyses, and found no difference between their merging
and non-merging sub-samples within the uncertainties. Thus, even among merging systems the Cheshire Cat appears to be hotter than expected based on scaling relations.

Using the four gravitational arcs that have measured redshifts and assuming the radius of
the arc is equal to the Einstein radius, we can estimate the mass inside the radius following, e.g., \citet{narayan96}:

\begin{equation}
M = \theta_e^2  \frac{c^2}{4G} \frac{D_s D_l}{D_{ls}},
\end{equation}
where $\theta_e$ is the angular Einstein radius, $c$ is the speed of light, $G$ is the gravitational constant, $D_s$ and $D_l$ are the
angular diameter distances to the lensed galaxy and the lens, respectively, and $D_{ls}$ is the distance between the lensed galaxy
and the lens. The four arcs have angular radii of 9\farcs0, 9\farcs7, 12\farcs3, and 12\farcs5
(51, 55, 70, and 71 kpc, respectively)
corresponding to lensed galaxies at redshifts 0.97, 2.78, 0.80, and 2.20, leading to mass estimates of
$2.3 \times 10^{13}$ \Msol, $1.8 \times 10^{13}$ \Msol, $5.3 \times 10^{13}$ \Msol, and $3.3 \times 10^{13}$ \Msol, respectively. These are comparable to
the values obtained by \citet{belokurov09} and \citet{kubo09} for the arc redshifts that were known at the time of those
studies. The uncertainties of the positions of the centers of the arcs and therefore the angular radii of the arcs of $\sim$$1\farcs5$ lead to uncertainties on the
mass estimates of 20\%--30\%, which may explain why the masses do not increase monotonically with radius. The above formula is also only strictly accurate for
a point source mass rather than an extended mass distribution.
Still, it is clear that 
the large lensing masses within a small volume demonstrate that this is a very mass-concentrated group, in agreement with the very high $c_{200}$ parameter of
17--34 as measured by \citet{bayliss11} and \citet{wiesner12}.

\subsection{A Nearly Certain Line of Sight Merger}
\label{ssec:merger}

The dynamical state of the Cheshire Cat is undoubtedly complex, but a number of results indicate that the system has
experienced a merger very recently. The most likely scenario is that the eastern eye
was the central galaxy in a group that collided with a roughly comparable mass
group centered on the western eye in a direction nearly along the line of sight. The most compelling evidence is the double-peaked
velocity distribution (Figure~\ref{fig:histo}), with $\sim$19 galaxies centered at 0.4278 and $\sim$29 galaxies centered
at 0.4329, a group rest frame velocity difference of 1350 km s$^{-1}$. It is unlikely that this velocity difference is due to the Hubble flow
with the two groups separated along the line of sight by a proper distance of 12 Mpc for three reasons. First, if these were two separate,
non-interacting groups each would have its own much fainter X-ray halo centered on its own BGG which would not have interacted yet
given the 12 Mpc separation.
% Or if the smaller (eastern) group lacked a detectable halo because of its smaller size, we would
%see a single X-ray halo centered on the larger (western) BGG.
Instead, we see only one X-ray halo that is not centered on either
bright elliptical eye, but whose centroid is located approximately midway between the two eye BGGs. The collision of the two X-ray halos
along the line of sight would most likely not produce any spatial asymmetries in the X-ray contours as is commonly seen in
plane-of-the-sky mergers, which explains why the X-ray surface brightness is well-fit by a single $\beta$-model.

Second, none of the gravitational lens arc centers coincide with either eye, but instead all cluster within 9 kpc of one another
at a position midway between the eyes near the position of the X-ray centroid (Figure~\ref{fig:circlefit}).
This is particularly relevant for the arc centers, which are
most sensitive to the dark matter distribution, indicating that neither eye is located at the center of the mass distribution
of the group. In addition, the fact that neither eye is even at the center of its own galaxy distribution in velocity space
(see Figure~\ref{fig:histo}) suggests that both galaxies have been dislodged from their respective dark matter distribution
centers during their gravitational interaction during the merging process.
% The fact that both galaxies have a smaller velocity
%than their respective sub-groups is expected in this situation. If the galaxies had a small impact parameter and swung around each other
%once the merger began, some of their line of sight velocity would be shifted to plane-of-the-sky velocity, leading to smaller radial velocity measurements,
%while at the same time displacing them from their sub-group dark matter centers in the plane-of-the-sky to their current projected separation distance.

Third, the temperature of the
X-ray gas in the central 20$^{\prime\prime}$ (110 kpc) is 5.4 keV and the global temperature of 4.3 keV is hotter than would be expected for 
a group of this mass or richness value. While the scaling relations of \citet{lopes09} were derived using largely normal groups rather than
specifically fossil groups, we can look to fossil groups of similar galaxy richness in the literature for a comparison. In the
fossil group sample of \citet{miller12}, the two fossil groups
with $N_{200}$ greater than 20 have measured global X-ray temperatures of 2.6 and 1.6 keV. The \citet{lopes09}
relation for $N_{200}$ versus X-ray luminosity predicts an X-ray luminosity of $1.4 \times 10^{43}$ erg s$^{-1}$ for the Cheshire Cat which is a factor of 4.5 below
what is measured, while the two $N_{200} > 20$ fossil groups of \citet{miller12} have soft X-ray luminosities $\le 1 \times 10^{43}$ erg s$^{-1}$,
in accordance with the \citet{lopes09} relation.
The elevated X-ray temperature and luminosity of the Cheshire Cat are characteristic of shock heating brought about
by a high velocity impact between the two groups. We can roughly estimate the Mach number of the collision from the ratio
of the estimated pre-shock temperature of the larger group prior to impact, 2.0 keV (the assumed pre-merger temperature of the G2 group),
to the post-shock temperature of 5.4 keV in
the central regions of the group. Applying the Rankine-Hugeniot equations with a temperature jump of a factor of $\sim$2.7
we estimate the Mach number $M \sim 2.4$ \citep[see][]{markevitch07}.
Given that the sound speed of an ideal gas at 2.0 keV is approximately 720 km s$^{-1}$ and assuming that the velocity difference between the
two sub-group peaks was 1350 km s$^{-1}$, a similar Mach number ($M \sim 1.9$) is obtained if we assume the collision has happened along or nearly along our
line of sight.

In summary, there is abundant evidence that two groups, with mass ratios of 1--2:1 have collided along our
line of sight. The resulting supersonic collision has shocked heated the gas in the center of the group to $>$5 keV, and significantly raised the
total X-ray luminosity of the gas as a result.

\subsection{The Past and Future of the Cheshire Cat and Its Implications on Fossil Groups}
\label{ssec:future}

It seems likely that the Cheshire Cat is composed of two separate groups that have recently begun the merging process. If true, then the 19
spectroscopically confirmed members of the eastern eye group (G1) might have composed a small fossil group prior to the merging, as discussed in
\S~\ref{sec:gemini}. It is likely that
the eastern eye that was the BGG of that group was more than two magnitudes brighter than the next brightest galaxy in that group (\S~\ref{sec:gemini}). It is
also likely this group had an X-ray halo exceeding $L_{X,bolometric} > 10^{42} h_{50}^{-2}$ erg s$^{-1}$ based on $N_{200}-L_X$ scaling relations
to satisfy the formal definition of a fossil group.
For our choice of $h$ and after converting from bolometric to 0.1--2.4 keV luminosity (assuming a 1--2 keV group temperature expected for an
$N_{200}=7$ group), this luminosity threshold becomes $4 \times 10^{41}$ erg s$^{-1}$. The \citet{lopes09} relation between $N_{200}$ and $L_X$ predicts
$L_X = 2 \times 10^{42}$ erg s$^{-1}$ for an $N_{200}=7$ group, with all four groups in the \citet{lopes09} sample with $N_{200}=6-10$ actually above the best-fit relation (all above 
$4 \times 10^{42}$ erg s$^{-1}$). Given that the expected $0.1-2.4$ keV luminosity of G1 is an order of magnitude above the  \citet{jones03} threshold, it is very likely
that G1 met the X-ray criterion of a fossil group before its merger with G2. Similar arguments make it likely that the G2 group also had a high enough X-ray luminosity to
fulfill the \citet{jones03} X-ray criterion, although it marginally failed both the $\Delta m_{12}$ and $\Delta m_{14}$ fossil group criteria.
%The remaining 29
%Cheshire Cat galaxies that composed the G2 prior to merging do not quite satisfy the definition of a fossil group, since the magnitude gap
%between tit BGG and the second rank galaxy is only 1.8 magnitudes. Still, this group
%was almost surely more luminous than $L_X > 4 \times 10^{41}$ erg s$^{-1}$ halo given its $N_{200}$ value of 15. Even if the two magnitude gap were fulfilled, the
%ongoing merging of the two nuclei (WA and WB in \S~\ref{sec:hst}) of the western eye of
%nearly equal brightness would preclude its classification as a fossil group.

We can estimate the time scale for the the `E' and `W' eye galaxies to merge into one system, creating an even more luminous galaxy at the center of the
Cheshire Cat.
%It is interesting to consider whether G2 would have become a fossil group on its own if it had not merged with G1.
%Given the small separation between the two nuclei WA and WB of 2.1 kpc ($0.37^{\prime\prime}$) these two nuclei should merge within the extended
%optical halo emission on a short time scale.
We utilize the results of \citet{kitzbichler08} who use a
semi-analytic model based on the Millennium $N$-body simulation to estimate the merging rate and time scale of galaxy pairs. They find that the
merging time scale is $T \approx 1.6 r_{\star} M_{\star}^{-0.3}$ Gyr for galaxies where the line of sight velocity difference between the two galaxies is less
than 3000 km s$^{-1}$, where $r_{\star}$ is the maximum projected separation of the two galaxies in units of 35.7 $h_{70}^{-1}$ kpc and $M_{\star}$ is the stellar mass
of the galaxies in units of $4.3 \times 10^{10} h_{70}^{-1}$ \Msol.
After the eye galaxies merge, they would have an apparent \sloanr~magnitude of 18.24, which once corrected for K-corrections and elliptical galaxy evolutionary
corrections \citep{roche09,girardi14} leads to an absolute magnitude of the merged galaxy of $M_r = -24.0$. Assuming
a mass-to-light ratio of six in the \sloanr~band for an elliptical galaxy, the approximate mass divided equally
between the two galaxies is $8.9 \times 10^{11}$ \Msol. This would give a merging time of $\sim$0.9 Gyr. A similar calculation indicates that the dual nucleus
of the western eye (WA and WB) should merge on a time scale of just a few tens of Myr, well before the eyes themselves merge.

Following the merger, the merged galaxy will be more than two magnitudes brighter than the next brightest spectroscopically confirmed galaxy within 0.5 $r_{200}$
of the combined system, which we estimate to be 0.5 Mpc for an eventually relaxed $N_{200} = 23$ group (\S~\ref{sec:mass}). Within this radius, there
are two galaxies without spectroscopic redshifts that are bright enough to ruin the fossil group status of the merged system by violating the
$\Delta m_{12} > 2.0$ criterion. However, these galaxies are both quite blue and are statistically unlikely to be located within the Cheshire Cat.
If the $\Delta m_{14} > 2.5$ criterion is used instead, the merged group will qualify as a fossil group regardless of the group membership status of the two blue galaxies.
The X-ray criterion of a fossil group should be easily satisfied.
Even after the high shock-induced X-ray luminosity declines, the expected X-ray luminosity for a relaxed $N_{200}=23$ group is $few \times 10^{43}$ erg s$^{-1}$
\citep{lopes09}, well above the \citet{jones03} X-ray criterion.

%The absolute magnitude of all components combined of the western eye is --22.7 in the observed \sloanr~band. Given that this roughly corresponds to
%the rest frame \sloang~band magnitude for a redshift $z=0.431$, if we assume a typical BGG \sloang -- \sloanr~color of 0.78 \citep{roche10}, the approximate \sloanr~
%absolute magnitude in the rest frame of the galaxy is --23.5. Assuming
%a mass-to-light ratio of six in the \sloanr~band for an elliptical galaxy, the approximate mass divided equally
%between the two galaxies is $5.7 \times 10^{11}$ \Msol. This would give a merging time of only 0.04 Gyr, after which the merged western eye would have
%been 1.8 magnitudes brighter than the next brightest galaxy in its sub-group (SDSS J103843.32+484904.8). Thus, if the western eye group had
%been allowed to evolve without merging with the eastern eye group, it would have nearly reached fossil group status in just a few tens of Myr.

High mass concentrations $c_{200}$ are expected of fossil groups (\citealp[e.g.][]{khosroshahi07}), so the high measured $c_{200}$ value already present if the
Cheshire Cat system is not unexpected given that both G1 and G2 were believed to be fossil and near-fossil groups prior to the merger. However,
in this situation, a high mass concentration is not necessarily evidence of the fossil group (or near-fossil group) natures of the initial subgroups.
Line of sight mergers are known to increase concentration parameters markedly (\citealp[e.g.][]{king07}). Thus, it is not possible to disentangle the
contribution to $c_{200}$ from the possible fossil group nature of the subgroups from the boost to $c_{200}$ given to the system from
the geometry of the merger.

It is interesting to note that that given the merger time scale, any observer who is currently within $z \sim 0.3$
of the Cheshire Cat should already see a massive fossil group with a luminous $M_r =$ --24.0 BGG at its center. Most fossil groups identified to date have
been at $z =0.3$ or closer. Thus, from our point of view the Cheshire Cat gives us
the opportunity to study the properties of a massive fossil group progenitor before its final assembly that can be compared to more nearby systems that
have already achieved fossil group status.

%The two eyes themselves should eventually merge. Given their current projected separation of 52 kpc, the above formula predicts that the
%merging timescale is $\sim$0.8 Gyr. At this time, the total \sloanr~rest frame absolute magnitude of the merged system will be
%--24.4, and this merged galaxy will easily be more than two magnitudes more luminous than the next brightest galaxy within 0.5 $r_{200}$). Furthermore,
%even after the high shock-induced X-ray luminosity declines, the expected X-ray luminosity for a relaxed $N_{200}=23$ group is $few \times 10^{43}$ erg s$^{-1}$
%\citep{lopes09}, well above the \citet{jones03} X-ray criterion for a fossil group. Given the merger time scale, any observer who is currently within $z \sim 0.30$
%of the Cheshire Cat should see a massive fossil group with a very luminous $M_r =$ --24.4 BGG at its center. From our point of view the Cheshire Cat gives us
%the opportunity to study the properties of a massive fossil group progenitor before its final assembly.

Despite not being a fossil group yet, our current view of the Cheshire Cat illustrates the ambiguity of fossil group definitions. The system G1 most likely was
already a fossil group, but had its fossil group status removed when it merged with G2. The group will transition back to fossil group status once the eye
galaxies merge in a Gyr, but the fossil group status is always subject to change if even one new luminous system enters the system via a future merger, and future
observers will classify the object as a normal group.
% The group
%is already heavily mass-concentrated (as evidenced by its $c = 17-34$ concentration parameter), which is believed typical of groups that were
%assembled cosmologically early, before formally becoming a fossil group.
%If at some later time after the eyes merge even one luminous elliptical galaxy falls into the group, the fossil group status will be removed, and it will be
%classified as a normal group by future observers, despite the 
%large mass concentration and
%underabundance of moderate luminosity ellipticals
In the local Universe, this appears to be happening to the NGC~1407 group, which would be classified as a fossil group if it were not
for the high velocity infall of NGC~1400 into the group at the moment \citep{su14}.
In this regard, the $\Delta m_{14}>2.5$ mag criterion for a fossil group may be more robust against the infall of a single luminous galaxy than the
$\Delta m_{12}>2.0$ mag criterion for a fossil group, but both methods are susceptible to the infall of a group with more than one moderately large elliptical galaxy.
Moving in and out of a fossil group phase is expected, as $N$-body simulations of the evolution of galaxy groups \citep[e.g.][]{vonbendabeckmann08,dariush10,cui11,gozaliasl14},
indicate that few fossil groups identified by either magnitude gap definition at high redshift survive to remain fossil groups at present, and few fossil groups
at $z=0$ were identified as fossil groups at high redshift.
Yet if there is a fundamental difference between fossil groups and normal groups, in the sense
that fossil groups assembled earlier as the name implies, and manifested by larger mass concentration parameters and X-ray scaling relations that
more closely resemble clusters than groups, such differences will not be
altered by the infall of a modest number of moderately luminous galaxies. This highlights the problems of defining an efficient method for identifying
truly old galaxy systems. Indeed,  \citet{dariush10} finds that neither the $\Delta m_{12}>2.0$ mag nor the $\Delta m_{14}>2.5$ criteria are particularly efficient at selecting
systems that formed early in their simulations.
%Alternatively, early-formed systems
%might be better identified as having $L_{BGG}/L_{total}$ (where $L_{BGG}$ is the optical luminosity of the BGG and $L_{total}$ is the total summed luminosities of all the group galaxies)
%above some threshold value regardless of magnitude gaps. However, when \citet{harrison12} calculated the fraction of the total group optical light that is contained in the
%central BGG they found that while the ratio for normal groups was below 50\% as expected, fossil groups exhibited a broad range of values between 25\%
%and 85\% (the two central galaxies of the Cheshire Cat combine for $\sim$50\% of the total optical light of the group). Thus, $L_{BGG}/L_{total}$ is also
%not an unambiguous indicator of fossil group status.
We note that the simulations do suggest, however, that the large mass ($1.2-1.5 \times 10^{14}$ M$_{\odot}$) and $\Delta m_{14}$ (2.70 mag)
of the future Cheshire Cat post-merger will allow the Cheshire Cat to
remain a fossil group longer than a typical fossil group. \citet{gozaliasl14} find that 40\% of fossil groups with masses exceeding $10^{14}$ M$_{\odot}$ (which might be more accurately called fossil clusters) retain
their fossil group status from $z=1$ to $z=0$, a much larger fraction than was found for lower mass fossil groups. Also, \citet[]{dariush10} find that fossil groups that meet the $\Delta m_{14}>2.5$ criterion remain fossil groups for an additional
Gyr on average compared to fossil groups that meet the $\Delta m_{12}>2.0$ criterion. So there is a reasonable chance that the Cheshire Cat will remain a fossil group for quite a while following the merger.

Previous studies have noted the lack of cool cores in some fossil groups \citep{khosroshahi04,khosroshahi06,sun04}, which is unexpected if fossil groups
are old systems that have been passively evolving since early times. However, if some fossil groups form from the merger of two fossil (or near-fossil) groups
like the Cheshire Cat system, any pre-existing cool core in one or both of the pre-merger groups will likely be destroyed by shock heating due to the merger.
Given the high relative velocity between G1 and G2, any cool core present within either group would have been destroyed by the merger. We calculated in
\S~\ref{sec:xray} that the central cooling time of the Cheshire Cat is 3 Gyr, and it will taken even longer for gas to cool out at a sufficiently large
radius for a cool core to be identified observationally at the redshift of the group. On the other hand, the group should satisfy the condition for a fossil
group once the two eye galaxies merge in about 0.9 Gyr. Thus, for at least 2 Gyr the Cheshire Cat will appear as a fossil group without a cool core.
This implies that if the merger of two fossil (or near-fossil) groups can be a common avenue for the formation of fossil groups, the lack of well-developed cool cores
in some present-day fossil groups is a consequence of the lag in time between the final galaxy merger and the establishment of a cool core following
the shock heating of the intragroup media of each merging sub-group.

%Indeed, the drawbacks of rigidly defining fossil groups as only those groups with a magnitude gap of 2.0 between the first and second
%rank galaxies has already been pointed out numerous times in the literature. Attempts to define fossil groups by magnitude gaps
%between the first and fourth ranked galaxies \citep[e.g.,][]{dariush10} rather than the first and second ranked galaxies might provide a more
%effective way of finding and classifying groups that formed at earlier cosmological epochs as fossil groups are believed to do. Alternatively, fossil groups
%could be identified as having a $L_{BGG}/L_{total}$ (where $L_{total}$) is the total stellar luminosity of all the group galaxies)
%above some threshold value regardless of magnitude gaps.
%Such a definition would allow for the occasional infall of a luminous galaxy without changing the status of the group as a fossil group.

\subsection{Similar Systems to the Cheshire Cat Group}
\label{ssec:comparison}

We searched the literature for descriptions of other groups of galaxies that could potentially be fossil group progenitors, i.e., systems with two or more giant elliptical
galaxies close enough to merge in less than a few Gyr surrounded by a collection of galaxies with brightnesses at least two magnitudes fainter than the
expected brightness of the final merged galaxy. J054-0309 studied by \citet{schirmer10} is
believed to be a fossil group falling into a spiral-dominated sparse cluster, although in this case there is only one bright elliptical. Also, the hot ICM is cooler than 
expected (rather than hotter than expected like the Cheshire Cat) with no evidence of shock heating indicative of a recent merger. There also is no clear evidence of a
double-peaked velocity distribution of the galaxies. In fact, \citet{schirmer10} speculate that rather than a sparse cluster, this component may be galaxies feeding an existing
fossil group along a filament. While interesting, this is substantially different from what we propose for the Cheshire Cat.

UGC842 \citep{lopesdeoliveira10} is a low-mass fossil group that exhibits a double-peaked velocity distribution, and a higher than expected ICM temperature indicative
of shock heating, similar to the Cheshire Cat, but with one velocity peak dominated by ellipticals and the other velocity peak containing a significant fraction of spirals. 
However, the entire system also contains only one giant elliptical galaxy, and \citet{lopesdeoliveira10} were unable to determine whether the two subgroups are
interacting or merely projected along the line of sight.

\citet{zarattini14} point out that the group FGS06 \citep{santos07} is composed of
two bright ellipticals of nearly equal brightness, and then a large gap between the second-rank galaxy and the third-rank galaxy. If the second-ranked galaxy were removed,
the system would fail the $\Delta m_{12}>2.0$ mag criterion for a fossil group, although it would pass the $\Delta m_{14}>2.5$ mag criterion.
\citet{zarattini14} suggest this is a
transitional fossil system. However, the two brightest galaxies are so far apart (projected separation of nearly 300 kpc) that they will not be merging any time
soon to create a fossil group like the Cheshire Cat. In other words, FGS06 is ``stuck" in this transitional phase for the foreseeable future.

Perhaps the closest analog to the Cheshire Cat is CL 0958+4702, whose constituent central galaxies will coalesce with the brightest cluster galaxy
in 110 Myr \citep{rines07}. While this combined galaxy will only be 1.3 mag brighter than the future second-ranked galaxy, a substantial amount of intracluster light (totaling
$10^{11} L_{\odot}$) surrounding the merging galaxies in a halo is expected to be accreted by the merged galaxy in a Gyr. If the entire amount of intracluster light
is accreted, the magnitude gap between the merged system and the second-ranked galaxy will be $\sim$2.0 mag, possibly qualifying this system as a fossil group progenitor.
The hot gas component of CL 0958+4702 is somewhat cooler and less X-ray luminous than the Cheshire Cat with a 50\% smaller virial mass, and does not show evidence
for merging in either its X-ray properties or galaxy velocity distribution as the Cheshire Cat does.

A lack of a substantial number of identified fossil progenitors does not necessarily indicate that the Cheshire Cat is rare or unique, but could result from the fact that a concerted
effort has not been undertaken to find such systems.
A search for more groups transitioning from the non-fossil phase to the fossil phase via mergers would lend insight into how fossil groups form. We have begun
an effort to identify galaxy systems with two or more bright elliptical galaxies surrounded by a population of much fainter galaxies, in which the brightest
elliptical galaxies are expected to merge on time scales of less than a few Gyr to form a fossil group by z$\sim$0.1 or less. By forming a sequence of systems with a variety of merging
time scales, we can form an evolutionary sequence of groups from non-fossil to fossil phase. Millennium simulations predict that $L^{\star}$ evolves differently as a function of redshift
for fossil groups than it does for non-fossil groups \citep{gozaliasl14}. A comparison of the galaxy luminosity function and $L^{\star}$ of progenitors
might indicate whether groups identified as fossil group progenitors more closely follow the $L^{\star}$ relation for fossil groups or non-fossil groups. The X-ray scaling relations of
fossil group progenitors might also differ from those of either fossil groups or non-fossil groups, particularly if shock heating due to merging subgroups as is believed to be happening
in the case in the Cheshire Cat is common.
To date, most simulations have focused mainly on the properties of groups once they have already reached the fossil phase, with little emphasis on the fossil group progenitor phase
other than how long the groups stay in this progenitor phase on their way in/out of this phase. It would be useful if future
simulations included a more detailed study of the fossil group progenitor phase as the group transistions from non-fossil group to fossil group, tracing such quantities as the galaxy
luminosity function of groups in the progenitor phase, shock heating due to heating and the subsequent cooling time scales of the hot gas, and galaxy merging time scales for the last
two large galaxies to become the BGG of the eventual fossil group.

\section{Summary}
\label{sec:summary}

Our optical and X-ray study of the Cheshire Cat gravitational lens system indicates that this system is one of the most likely (and potentially the first)
example of a fossil group progenitor.
Our primary findings are:

1) We obtained {\it Gemini} GMOS-determined spectroscopic redshifts for 34 group galaxies, and added 14 more from the literature. The group shows a bimodal velocity
distribution, with peaks separated by 1350 km s$^{-1}$. The two eye galaxies lie in different peaks, although neither eye is at the center of
its respective velocity peak. Using our imaging and velocity information along with published galaxy richness and X-ray scaling relations, we find that for the Cheshire Cat
$N_{200} = 23$, $\sigma_{los} = 659 \pm 69$ km s$^{-1}$, $r_{200}$ = 1.0--1.3 Mpc, and $M_{200} = (1.2 - 1.5) \times 10^{14}$ \Msol.
Dividing the galaxies into two separate
sub-group distributions gave velocity dispersions of $\sigma_{los} = 318 \pm 51$ km s$^{-1}$ and $\sigma_{los} = 337 \pm 42$ km s$^{-1}$, and virial masses of
$(0.33 \pm 0.07) \times 10^{14}$ \Msol~ and
$(0.39 \pm 0.06) \times 10^{14}$ \Msol, respectively, for the eastern eye
group (G1) and the western eye group (G2). On the other hand, group scaling relations predict that G2 is approximately twice as massive as
G1.  Four separate gravitational arcs
give mass estimates of $few \times 10^{13}$ \Msol~inside a radius of $\sim$60 kpc.

2) The Cheshire Cat shows abundant evidence for merging of two groups of approximately equal mass along the line of sight.
In addition to a bimodal velocity distribution, the central X-ray temperature
and X-ray luminosity are much higher than expected for its galaxy richness even in comparison to fossil groups, indicating heating from an
$M \sim$ 1.9--2.4 shock. Also, the center of the
X-ray halo and of the four gravitational arcs are not located at the position of either eye galaxy, but midway between them, indicating that
neither of the two dominant galaxies in the system reside at the center of the mass distribution of the group.
Before the collision, the eastern eye was mostly likely the BGG of a small fossil group, and the western eye was most likely the BGG of a near-fossil group.

3) {\it HST} imaging reveals that the western eye is composed of two nuclei of similar brightness embedded off-center in an extended
stellar halo. The nuclei are separated by a projected distance of only 2.1 kpc, and should merge together in a few tens of Myr.
The two eye galaxies should merge in less than a Gyr to form a massive fossil group with a $M_r = -24.0$ BGG at its center, after which the system
will be a fossil group. Thus, the Cheshire Cat is a prime candidate for a fossil group progenitor. Observers currently within $z \sim 0.3$ most likely already see
the Cheshire Cat as a massive fossil group.

4) The time scale for the eyes to merge and for the system to satisfy the criteria of a fossil group is shorter than the estimated cooling time
of the hot gas in the core of the group. This means there will be a period of time where the system will be a fossil group without a cool core.
If the merging of fossil (or near-fossil) groups is a viable mechanism for creating a substantial fraction of present-day fossil groups, the observed
lack of cool cores in some fossil groups is understandable.

%4) The high mass concentration of the group indicates that it is already showing the characteristics of a fossil group despite not yet formally
%meeting the requirements of a fossil group.
%4)  We discuss the shortcomings of the current fossil group definitions, and the need
%to develop a more flexible definition for a fossil group that is not reliant on whether or not a single moderate luminosity galaxy falls into the
%group.

\acknowledgments Acknowledgments: 
We thank Rob Proctor, Ka-Wah Wong, Dacheng Lin, and Li Ji for useful conversations and feedback, and an anonymous referee for a thorough reading 
of the manuscript. We also thank Matthew Bayliss in assisting us in obtaining galaxy redshifts from his sample.
This work was supported by {\it Chandra} Grant GO0-11148X, and
based partially on observations obtained at the {\it Gemini} Observatory, which
is operated by the 
Association of Universities for Research in Astronomy, Inc., under a
cooperative 
agreement with the NSF on behalf of the {\it Gemini} partnership: the National
Science 
Foundation (United States), the National Research Council (Canada),
CONICYT (Chile), the
Australian Research Council (Australia), Minist\'erio da Ci\^encia, Tecnologia
e 
Inova\c{c}\~ao (Brazil) and Ministerio de Ciencia, Tecnolog\'ia e Innovaci\'on
Productiva 
(Argentina).

\bibliographystyle{apj}
\bibliography{apj-jour,preprint}

\end{document}